\documentclass[aps,twocolumn,pra,superscript,floatfix,superscriptaddress,showpacs,footinbib]{revtex4-1}
\usepackage{amssymb}

\usepackage[pdftex]{graphicx}
\usepackage{dcolumn}
\usepackage{bm}
\usepackage{amsmath}
\usepackage{array}
\usepackage{color}
\usepackage{float}
\usepackage{subfigure}
\usepackage{dsfont}
\usepackage{xcolor}

\newcommand{\+}{\dagger}

\newcommand{\e}{\varepsilon}

\newcommand{\leads}{\mathrm{lead}}

\newcommand{\qdot}{\mathrm{dot}}
\newcommand{\dotleads}{\mathrm{dot-lead}}
\newcommand{\dotM}{\mathrm{dot-M}}

\newcommand{\s}{\sigma}

\newcommand{\up}{\uparrow}
\newcommand{\down}{\downarrow}

\newcommand{\veck}{\mathbf{k}}

\usepackage{hyperref}
\hypersetup{
    colorlinks,%
    citecolor=blue,%
    linkcolor=blue,%
    urlcolor=blue
}

\begin{document}

\title{Robustness of the Kondo effect in a quantum dot coupled to Majorana zero modes}

\author{Joelson F. Silva}
\affiliation{Instituto de F\'{\i}sica, Universidade Federal de Uberl\^andia, 
Uberl\^andia, Minas Gerais 38400-902, Brazil.}

\author{Luis G.~G.~V. Dias da Silva}
\affiliation{Instituto de F\'{\i}sica, Universidade de S\~{a}o Paulo,
Rua do Matao 1371,  S\~{a}o Paulo, SP  05508-090,  Brazil}

\author{E. Vernek}
\affiliation{Instituto de F\'{\i}sica, Universidade Federal de Uberl\^andia, 
Uberl\^andia, Minas Gerais 38400-902, Brazil.}
\affiliation{Department of Physics and Astronomy, and Nanoscale and Quantum Phenomena Institute,
Ohio University, Athens, Ohio 45701--2979, USA}
%


\begin{abstract}
The prospect of using semiconductor quantum dots as an experimental tool to distinguish Majorana zero modes (MZMs) from other zero-energy excitations such as Kondo resonances has brought up the fundamental question of whether topological superconductivity and the Kondo effect can coexist in these systems.
Here, we study the Kondo effect in a  quantum dot coupled to a metallic contact and to a pair of MZMs. We consider a situation in which the MZMs are spin polarized in opposite directions. By using numerical renormalization-group calculations and scaling analysis of the renormalization group equations, we show that the Kondo effect takes place at low temperatures, regardless the coupling to the MZMs. Interestingly, we find that the Kondo singlet essentially decouples from the MZMs such that the residual impurity entropy can show local non-Fermi liquid properties characteristic of the single Majorana excitations. This offers the possibility of tuning between Fermi-liquid and non-Fermi-liquid regimes simply by changing the quantum dot-MZM couplings. 
 
\end{abstract}
\maketitle

\section{Introduction}

Majorana zero modes (MZMs) are known to emerge as a low-energy excitations in a variety of condensed matter systems \cite{Kitaev,Liu,Alicea-1,Read}. The simplest and perhaps the most experimentally investigated are topological superconducting quantum wires (TSQWs), in which the MZMs appear bound to the edges of the wires \cite{Oreg,Nayak,Mourik,Deng,Das} when the experimental  parameters are tuned to the topological regime. The proposed experimental setups to realize TSQWs involves the use of quantum wires made of materials with strong spin-orbit coupling in proximity with an s-wave superconductor~\cite{Sau2,Oreg}. External magnetic fields are applied to break time-reversal symmetry (TRS) as to produce effectively spinless electrons with p-wave superconducting pairing, leading to the effective realization of the 1D Kitaev model~\cite{Alicea2}.

An interesting development is the coupling of quantum dots (QDs) to one of the edges of a TSQW\cite{PhysRevB.84.201308,Leijnse_2014,UEDA2014182,Liu,PhysRevB.98.075142,PhysRevB.96.045135,PhysRevB.94.155417,Silva_2016,Ramos_Andrade_2018}. It has been shown that the MZM ``leaks'' into the quantum dot, a phenomena that can be possibly observed by spectroscopy transport measurements by coupling the QD to source and drain leads~\cite{vernek1,Deng:Phys.Rev.B:98:085125:2018}. Another important effect arising due to the strong repulsive Coulomb interaction in a QD coupled to metallic leads is the Kondo screening of the QD effective magnetic moment, which governs low energy physics and inevitably occurs at low temperatures~\cite{Hewson}.  

Apart from some recent proposals with global time-reversal symmetric systems~\cite{Mele,Wang,Frustaglia}, it is well-established that TRS breaking is an important ingredient in the formation of TSQWs \cite{Read,Kitaev,Sau2,Alicea2,Fugimoto}. TRS breaking is, however detrimental for the Kondo effect. Moreover, superconducting pairing induced in QDs is known to compete with the Kondo screening effect \cite{Pillet}. As a result, both superconducting  pairing and TRS breaking induced in the QD by the TSQW may destroy the  Kondo effect, as discussed in earlier works \cite{Pillet}. It is interesting, however that recent theoretical studies suggest that the Kondo effect  coexists with MZMs in a QD-TSQW junction~\cite{Liu,PhysRevX.4.031051,Lee,Chirla:Phys.Rev.B:90:195108:2014,Tijerina,Gorski2018}, even the presence of ferromagnetic contacts~\cite{PhysRevB.95.155427}. Indeed, some of us have shown this coexistence in interacting QDs coupled to a metallic lead and to a single Majorana mode~\cite{Tijerina,Cifuentes}. However, the question of \textit{why} Kondo screening still takes place in the presence of MZMs is still not well understood.

\begin{figure}[t!]
\centering
\subfigure{\includegraphics[clip,width=3.5in]{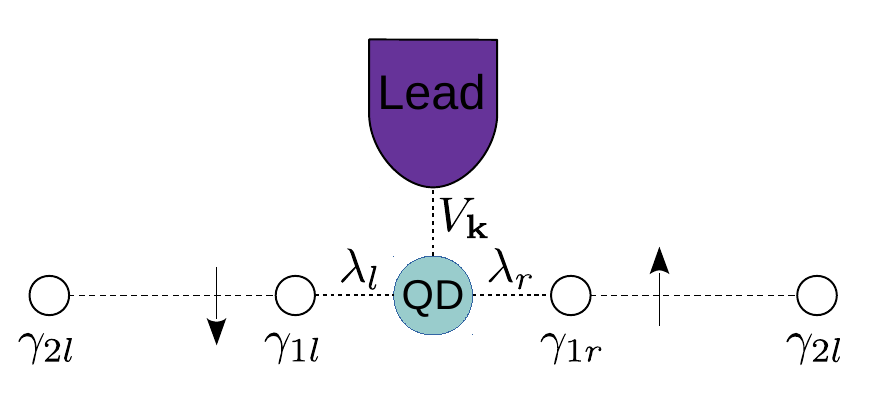}}
\caption{(Color online) Schematic representation of the model. The quantum dot (QD) couples to the  lead via matrix element $V_{\bf K}$ and to left (l) and right (r) Majorana modes, assumed to present opposite  polarization as denoted by the up and down arrows. } 
\label{model}
\end{figure}

In this work, we revisit the Majorana-Kondo problem in a QD coupled to a metallic lead and to a TSQW. We present a  detailed  investigation focusing on the mechanisms that make the Kondo effect resilient to the TRS breaking and to the superconducting pairing induced by the TSQW.  To this end, we consider a more general case in which the QD is coupled to two MZMs and to metallic leads~\cite{Adonai:Thesis:2016}. Moreover, we assume that  the MZMs are spin-polarized in opposite directions. This configuration can be obtained by coupling the QD to two spatially separated TSQWs, each of which is in proximity to a ferromagnet with opposite polarizations \cite{Hoffman:Phys.Rev.B:045440:2017}. In this configuration, the QD-MZM spin-dependent couplings arising from the wave-function overlap between dot states and the edge MZM states \cite{Prada:Phys.Rev.B:96:085418:2017} will only couple each spatially-separated MZM with the QD electronic states with the same spin direction, which we take to be ``up'' or ``down''. As a result, if the QD is symmetrically  coupled to both MZMs, local TRS in the QD is restored.

We consider and effective  single impurity Anderson model (SIAM) that describes the coupling of the ``impurity'' (QD and MZMs)  with a metallic lead. The low-energy properties of this model are investigated using Wilson's NRG method as well as Anderson poor's man scaling. The NRG calculations allow for a quantitative description of local properties at the fixed points of the model, particularly the impurity contribution to the entropy at a given temperature $S_{\rm imp}(T)$ . The results show that the low-energy limit $S_{\rm imp}(T \rightarrow 0)$ assumes Fermi-liquid-like values $\ln{2}$ (or $2\ln{2}$) if both (or none) MZMs are coupled to the QD. When only one of the MZMs is coupled, then $S_{\rm imp}(T \rightarrow 0) \!=\! (3/2)\ln{2}$, indicating a non-Fermi-liquid behavior marked by the presence of three free MZMs. More importantly, we establish that electron-electron interactions or the presence of Kondo correlations do not change this overall picture, as if the Kondo effect plays a ``spectator role'' in the process.

In order to understand the relevance of Kondo correlations to the flow to the low-energy fixed point, we derive an effective Kondo-like Hamiltonian to investigate how the coupling to both MZMs modify the Kondo physics of the model. This effective Kondo model is then studied within the traditional Anderson poor's man scaling renormalization analysis, that allows us to identify the evolution of the effective parameters under the renormalization procedure. We obtain a set of differential equations ($\beta$-function) for the effective coupling that can be solved numerically. 

As one of the central results of the paper, we show that the scaling equation for the effective coupling, $J$, associated to the Kondo effect appears fully decoupled from the other scaling equations,clearly indicating that the Kondo strong coupling fixed point remains intact in the presence of the TSQW. As a result, the Kondo temperature $T_K$ extracted from the solution for $J$ is insensitive to the parameters involving the TSQW. This result sheds light as to \emph{why} the $T_K$ obtained by the thermodynamic properties of the system calculated with the numerical renormalization group depends weakly with the coupling between the QD and a single MZM, as studied previously.

This paper is organized as follows. In Sec.~\ref{sec:model} we present the quantum impurity model describing the QD coupled to MZMs and a metallic lead and we present the NRG calculations for this model in Sec.~\ref{sec:NRG}. In Sec.~\ref{Poor_man_scaling} we derive a low-energy effective model and perform a perturbative scaling  analysis of the effective coupling parameters which nicely complements the NRG results. Finally, a summary of our work is presented in Sec.~\ref{sec:conclusions}.

\section{Model}\label{sec:model}
For concreteness, we consider an interacting QD coupled to a metallic lead and two topological superconductors. The topological superconductors as assumed to sustain  Majorana zero modes at their edges with different spins polarizations as schematically shown in Fig.~\ref{model}. The system is described by the following Anderson-like impurity Hamiltonian
\begin{eqnarray}\label{eq:Htot}
 H=H_{\qdot}+H_{\leads}+H_{\dotleads}+H_{\dotM},
\end{eqnarray}
in which
\begin{eqnarray}\label{H_dot}
 H_{\qdot}=\sum_{\sigma}\e_{d}d^\+_{\sigma}d_{\sigma}+Un_{d\uparrow}n_{d\downarrow},
\end{eqnarray}
describes the isolated quantum dot, where $d^\dagger_\sigma$ and $d_\sigma$ ate the operators the crates and annihilates an electron with energy $\e_d$ and spin $\sigma$ in the single level QD,  $n_{\sigma}=d^\dagger_\sigma d_\sigma$ is the number operator and $U$ represents the onsite Coulomb interaction at the dot.
\begin{eqnarray}
H_{\leads}=\sum_{ \veck,\sigma}\e_{\veck, \sigma}c^\+_{ \veck, \sigma}c_{ 
\veck, \sigma},
\end{eqnarray}
describes the  normal leads, with $c^\dagger_\sigma$ ($c_{\veck\sigma}$) being the operator that creates (annihilates) an electrons with momentum $\veck$, energy $\e_\veck$ and spin $\sigma$ in the normal metal. 
\begin{eqnarray}\label{Hdotleads}
H_\dotleads=\sum_{\veck, \sigma}\left(V_{\veck}d^\+_{\sigma}c_{ 
\veck, \sigma}+V_{\veck}^{*}c^\+_{\veck,\sigma}d_{\sigma}\right),
\end{eqnarray}
connects the dot to the normal leads via matrix element $V_\veck$. Finally,

\begin{eqnarray}\label{eq:H_T}
H_{\dotM}&=&i\lambda_{r}(e^{-i\phi_{r}2}d^\+_{\up}+e^{i\phi_{r}/2}d_{\up})\gamma_{r} \nonumber \\
&&\quad +\lambda_{l}(e^{i\phi_{l}2}d^\+_{\down}-e^{-i\phi_{l}/2}d_{\down})\gamma_{l}, \nonumber\\
\end{eqnarray}
describes the coupling between the QD and the Majorana zero modes in the topological superconductors edges. Here, the operators $\gamma_{l,r}$ are Majorana operators, with the property
\begin{equation}
 \gamma_{i}=\gamma_{i}^{\dag},
\end{equation}
and obeying the fermion anti-commutation relation 
$ [\gamma_{i},\gamma_{j,}]_{+}=\delta_{i,j}$
and $\phi_{r/l}$ represents the phase of the left/right topological superconductor. For convenience, we perform a gauge transformation $d_{\s}\rightarrow d_{\s}e^{-i\phi_{l}/2}$, upon which the expression \eqref{Hdotleads}  can be rewritten as
\begin{eqnarray}\label{eq:H_TG}
H_{\dotM}=i\lambda_{r}(e^{-i\delta\phi/2}d^\+_{\up}+e^{i\delta\phi/2}d_{\up})\gamma_{r}+
\lambda_{l}(d^\+_{\down}-d_{\down})\gamma_{l}, \nonumber\\
\end{eqnarray}
where $\delta\phi=\phi_{r}-\phi_{l}$ is the superconductors phase difference between the two superconductors. To address the Kondo physics, in the next section we perform NRG calculations which will allow us to understand the low-energy physics of the system.

We emphasize that the QD-MZM coupling strengths $\lambda_{l,r}$ in Eqs.~\eqref{eq:H_T} and \eqref{eq:H_TG} are, in general, spin-dependent and can couple to both dot spins depending on the respective ``spin canting angle'' $\theta_{l,r}$ as $\left(\lambda_{\up (l,r)},\lambda_{\down (l,r)} \right) \equiv \lambda_{l,r} \left(\sin{\frac{\theta_{l,r}}{2}}, -\cos{\frac{\theta_{l,r}}{2}}\right)$ \cite{Prada:Phys.Rev.B:96:085418:2017,Hoffman:Phys.Rev.B:045440:2017,Deng:Phys.Rev.B:98:085125:2018}. In this work, we take $\theta_r \! = \!\pi$ and $\theta_l \! = \! 2 \pi$ such that only a single dot spin operator is coupled to each MZM, making $H_{\dotM}$ fully spin-conserving. This choice adds an extra symmetry (spin parity $P_{\sigma} = (-1)^{N_{\sigma}})$ to the full Hamiltonian, which is important for the NRG calculations presented in Sec.\ \ref{sec:NRG}.

\section{Numerical Renormalization Group Analysis}
\label{sec:NRG}
\noindent 

In order to implement the Numerical Renormalization Group~(NRG) calculations, the first step is to write Hamiltonian \eqref{eq:Htot} in terms of Dirac operators. To this end, we combine two Majorana operators to define the conventional fermion operators as $f_{\up}=(\gamma_{1r} +i\gamma_{2r})/\sqrt{2}$ and $f_{\down}=(\gamma_{1l}+i\gamma_{2l})/\sqrt{2}$. As mentioned before, only 
the $\gamma_{1r}\equiv \gamma_{r}$ and $\gamma_{1l}\equiv \gamma_{l}$ modes are coupled to the QD. In term of this $f$-fermions, the Hamiltonian $H_{\dotM}$~\eqref{eq:H_TG} can be written as
\begin{eqnarray}\label{H_f}
 H_{\dotM}&=&\frac{i\lambda_{r}}{\sqrt{2}}[e^{-i\delta\phi/2}(d^\+_{\up}f_{\up} +d^\+_{\up}f_{\up}^\+)+
 e^{i\delta\phi/2}(d_{\up}f_{\up} +d_{\up}f_{\up}^\+)]\nonumber \\
 & &+\frac{\lambda_{l}}{\sqrt{2}}(d^\+_{\down}f_{\down}+d^\+_{\down}f_{\down}^\+ -d_{\down}f_{\down}-d_{\down}f_{\down}^\+).
\end{eqnarray}
We can now construct a Fock space for the occupation numbers of the QD electrons and of the $f$-fermions as  $\{|n_{d\sigma}\rangle \otimes |n_{f\sigma}\rangle \}$, where $n_{d\sigma}= d_{\sigma}^{\dag}d_{\sigma}$ and $n_{f\sigma}=f_{\sigma}^{\dag}f_{\sigma}$ with 
$\sigma=\up,\down$. Throughout the paper, we consider the wide-band limit for the (particle-hole symmetric) conduction band of the metallic lead. As such, the density of states of the metallic electrons is taken to be a constant, given by  $\rho_0=(2D)^{-1}$ for energies in the range $\omega \in [-D,D]$ and zero otherwise. In the following, we set the the Fermi energy at $\omega \!=\! 0$ and the half-bandwidth $D$ as our energy unit.

The presence of Majorana modes breaks gauge invariance in the system and introduces some technical difficulties for the NRG implementation as the total charge $N_{\rm occ} \equiv N_{\up} + N_{\down}$ with $N_{\sigma} = n_{d\sigma} + n_{f\sigma}$ is no longer a good quantum number \cite{Lee,Tijerina,Cifuentes}. Nevertheless, the parities for each spin $\sigma$, defined as $P_{\sigma} \!  \equiv \! (-1)^{N_{\sigma}}$, can be used as a good quantum numbers in our case, thereby reducing the block size of the Hamiltonians generated along the NRG procedure.

\begin{figure}[h]
\centering
\subfigure{\includegraphics[clip,width=3.5in]{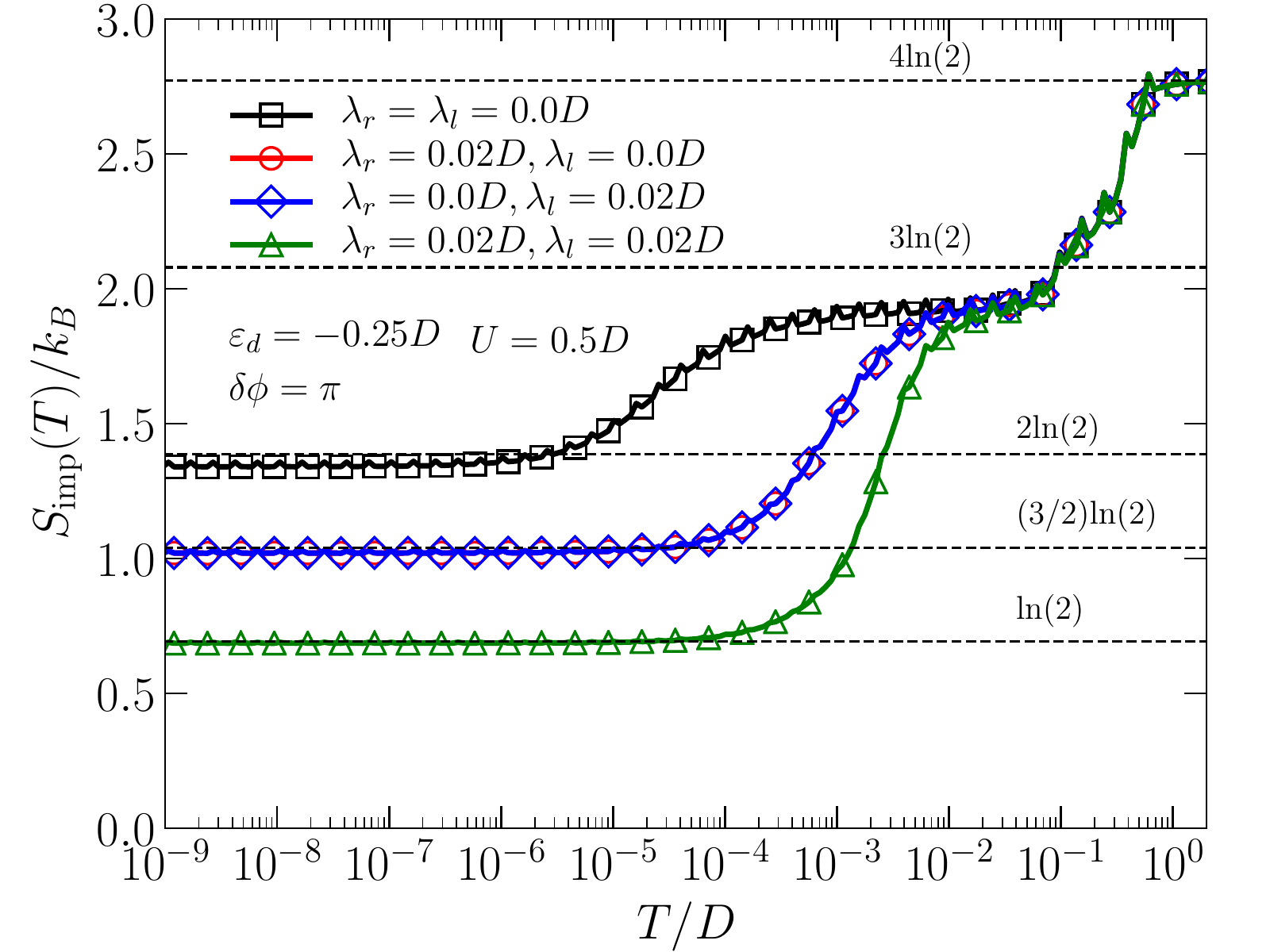}}
\caption{(Color online)~Impurity entropy extracted using NRG. In the presence of one Majorana mode the Kondo singlet and the local Majorana singlet coexists and the entropy presents the peculiar value $(3/2)\rm ln(2)$ for $T=0$.
} 
\label{entropy_NRG}
\end{figure} 

To get a better insight of the low-energy physics of the system, we use the NRG to calculate the contribution from the QD-Majorana system to the entropy. In the remainder of the paper, we  will refer to this quantity as the ``impurity entropy'' $S_{\rm imp}$. In the following, we use the same parameters as those in Fig~\ref{couplings-1} unless otherwise stated. 

Figure \ref{entropy_NRG} shows the impurity entropy as a function of temperature ($S_{\rm imp}(T)$) for different values of $\lambda_r$ and $\lambda_l$.  The different plateaus correspond to the fixed points of the renormalization flow of the  MZM-QD-leads system. The first plateau (large $T$) at $S_{\rm imp} \sim 4 k_B \ln{(2)}$ corresponds to the free orbital fixed point. As the temperature lowers, the system approaches the local moment fixed point, marked by a plateau at $S_{\rm imp} \sim 3 k_B \ln{(2)}$. Finally, for  $T\rightarrow 0$, the system approaches the strong coupling fixed point. The value of $S_{\rm imp}$ in this fixed point depends strongly on how many MZMs are directly coupled to the QD.

For $\lambda_{l}=\lambda_{r} = 0$ (black curve),  the results show that $S_{\rm imp}(T \rightarrow 0) \sim 2 k_B\ln(2)$. One can understand this result qualitatively as follows: in the strong coupling fixed point, the   QD is Kondo-screened in a singlet state, while the decoupled Majorana modes provide an additional double degeneracy to the ground state, yielding a $\ln{(4)}$ residual entropy. In fact, within the $f$-fermion representation, $f^{\dagger}_{\sigma}f_{\sigma}|n_{f, \sigma}\rangle = n_{f, \sigma}|n_{f, \sigma}\rangle$, we find four different possible zero-energy states which account for the four-fold ground state degeneracy. 

As we turn on the couplings to the Majorana modes, some of these degeneracies are lifted and the entropy goes to a lower value as $T \rightarrow 0$. The most interesting case occurs when only a single Majorana mode is coupled to the QD, i.e., $\lambda_{l}=0$, $\lambda_{r} \neq 0$ or vice-versa [circles (red) and diamond (blue) curves of Fig.~\ref{entropy_NRG}]. Notice that, in this case, the strong coupling fixed point plateau behaves as $S_{\rm imp}(T\rightarrow 0) \sim (3/2) k_B\ln(2)$. This is consistent with the fact that  there are now three decoupled MZMs at low energies and the MZM directly coupled to the QD does not contribute to the entropy.

When both couplings are nonzero (e.g., $\lambda_{r}=\lambda_{l}=0.02$ corresponding to the green curve with triangle symbols in Fig.\ \ref{entropy_NRG}), the ground state exhibits a residual entropy $S_{\rm res} \equiv S_{\rm imp}(T \rightarrow 0)=k_B\ln(2)$, stemming from the two MZMs that remain decoupled from the rest of the system. From these results, we can see that the low-energy residual entropy $S_{\rm res}$ takes the form  $S_{\rm res}=(N_0/2) k_B\ln(2)$, where $N_0$ is the number of uncoupled (``free'') Majorana modes. In fact, this result can be rigorously proved in the case of free MZMs (see Appendix \ref{sec:analytic}). Note that if $N_0$ is odd, the entire system behaves as a non-Fermi liquid. This is quite similar to the results obtained in the (unstable) non-Fermi-liquid fixed point of the two-channel Kondo problem, where in the Majorana representation there is a free Majorana mode left as $T\rightarrow 0$ 
\cite{PhysRevB.46.10812,PhysRevB.52.6611,Cox,ZHANG}. The fractional aspect of the entropy of a single MZM coupled to a quantum dot has also been subject of more recent studies \cite{Smirnov:Phys.Rev.B:92:195312:2015,PhysRevLett.123.147702}.

\begin{figure}[t]
\centering
\subfigure{\includegraphics[clip,width=3.5in]{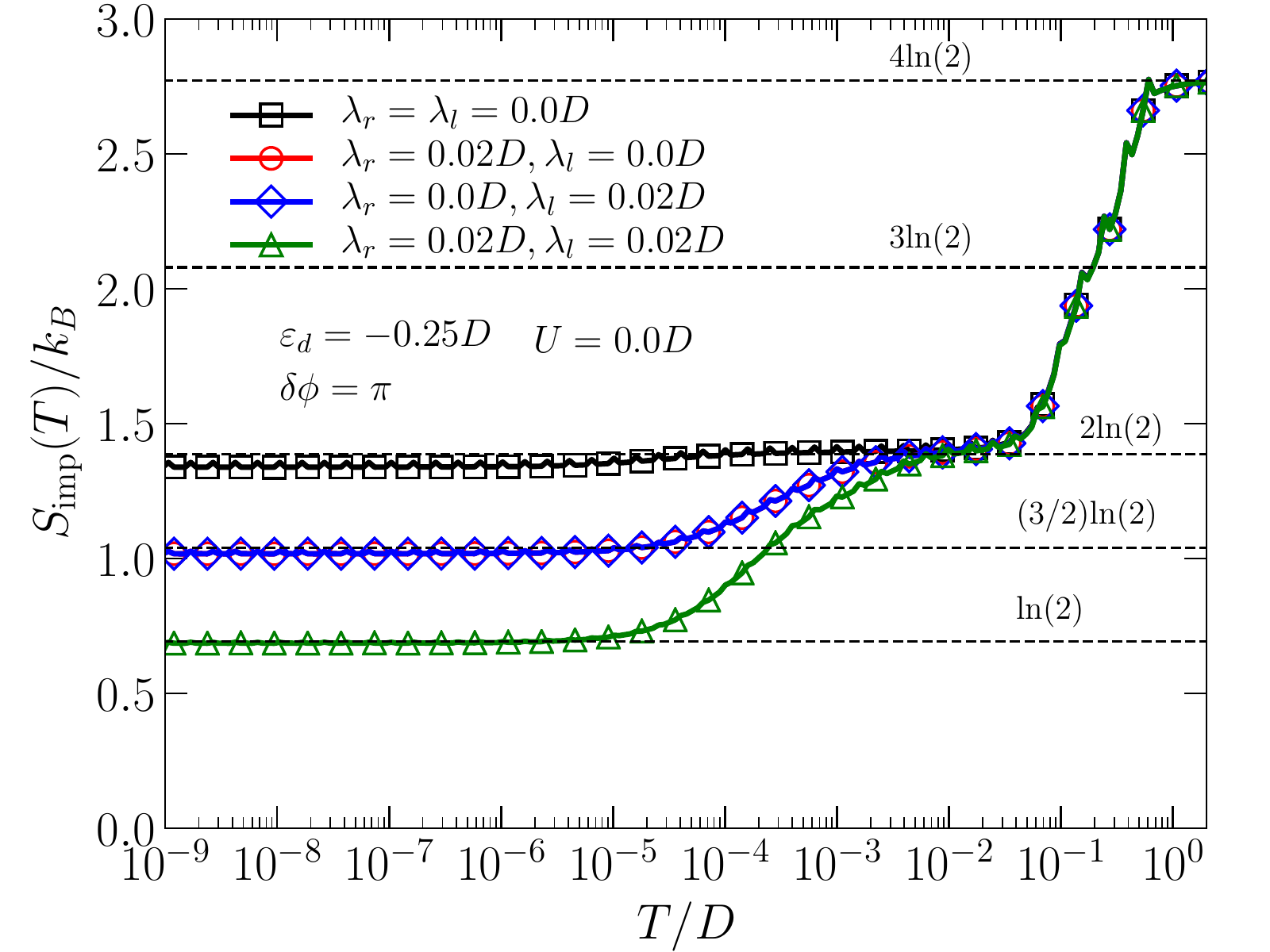}}
\caption{(Color online)~Impurity entropy extracted using NRG for the non-interacting regime~($U=0$).  For $T\rightarrow 0$ the entropy behaves as in the interacting case giving the same results  for $T=0$ which confirms that the residual entropy behavior is due only the free Majorana modes. } \label{entropy_NRG_U}
\end{figure}

To show that the residual entropy of the QD is dominated by the free Majorana modes of the system, and thereby is not affected by the Kondo singlet, it is interesting to consider the  the noninteracting regime. This can be accomplished by taking $U\!=\!0$ in \eqref{H_dot}. 
In this regime, no Kondo effect takes place and the entropy features are entirely due the coupling of the QD with Majorana modes and the metallic leads. As before, we compute  $S_{\rm imp}(T)$ with NRG. The results are shown in 
Fig~\ref{entropy_NRG_U} and the residual entropy follows precisely the behavior of the interacting case shown in Fig.\ \ref{entropy_NRG}. This result is also consistent with analytical calculations for $S_{\rm imp}(T \rightarrow 0)$ carried out for $U\!=\!0$ shown in Appendix~\ref{appendix_B}.

The impurity entropy in the noninteracting case (Fig.\ \ref{entropy_NRG_U}) displays a free orbital plateau ($S_{\rm imp} \sim 4 k_B\ln(2)$) at large temperatures, but not 
the local moment plateau ($S_{\rm imp} \sim 3 k_B\ln(2)$). This is expected as the local moment regime is not present for $U=0$. The interesting region occurs for $T\rightarrow 0$. In this regime, we have $S_{\rm imp}(T\rightarrow 0)$ displaying the same behavior as in the interacting case (Fig.\ \ref{entropy_NRG}). This indicates that, in the presence or absence of the Kondo singlet, the low-temperature behavior of the entropy is entirely determined by the free Majorana modes of the system. The absence of an additional contribution for the entropy  in the interacting regime shows that the Kondo singlet is preserved even in the presence of the Majorana modes, indicating a disassociation of the processes leading to Kondo screening at low-energies from the QD-MZMs couplings at high-energies. In order to clarity the origin of this interesting behavior, in the following we derive an effective Hamiltonian that describes the low-energy regime of the system and perform a RG scaling analysis.

To close this section, let us  briefly discuss the fractional residual entropy observed in the NRG results~(Figs.~\ref{entropy_NRG} and \ref{entropy_NRG_U}). The situation here is closely related to the non-Fermi liquid regime of the two-channel Kondo problem. As showed by Emery and Kivelson \cite{PhysRevB.46.10812}, the  bosonization~(refermionalization) approach to the two-channel Kondo problem uses a Majorana representation to map the problem into a  zero-energy resonant level coupled to metallic channels. In fact, only a ``half fermion''~(Majorana mode) in the resonant level is strongly coupled to the metallic electrons while the other ``half"  is free. This  ``free'' Majorana mode sets the non-zero value of the residual entropy, resulting in the celebrated result 
%
$S_{\rm imp}(T \rightarrow 0) \!=\! k_{\rm B} \rm ln(2)/2$.
%

This result reveals a special \emph{non-integer} ground state degeneracy, which is a characteristic of the non-Fermi liquid behavior of the two-channel Kondo problem~\cite{Nozieres}. In fact, Affleck and Ludwig, using a boundary conformal field approach, showed that some quantum critical systems (including the  the multichannel Kondo impurity problem), may display a non-integer ground state degeneracy~$g$, whose specific value of which depends on the universality class of the boundary conditions~\cite{PhysRevLett.67.161}.

In the present context, the system (namely two topological quantum wires coupled to an interacting QD) is distinct  from the two-channel Kondo problem. However, as it has already been pointed out by some of us in previous works~\cite{vernek1,Tijerina}, a Majorana mode from the edge of a wire can ``leak'' into the QD once it is tunnel coupled to it. The scenario of MZMs coupled to the QD~(and to a metallic lead) and their correspondent free partners in the other edge~(see Fig.\ref{model}) can be viewed as a formal analogue of the two-channel Kondo problem, but with a different physical origin. A detailed calculation of the impurity residual entropy and the connection with the effective model are shown in the appendices~\ref{sec:analytic} and \ref{appendix_B}.

We should also point out that such fractional character of the residual entropy may be used to distinguish MZMs from trivial zero-energy states in experiments. Indeed, a protocol based in entropy measurements to detect MZMs was recently  proposed by Sela {\it et al.}~\cite{PhysRevLett.123.147702}. Their proposal consists of a MZM coupled to a metallic lead close to a QD in the Coulomb blockage regime. The QD occupation can change from $N$ to $N+1$ by changing the chemical potential $\mu$ of a reservoir coupled to the QD. This change can be monitored by a charge detector and it is related to the local entropy via the Maxwell's relation $(dS/d\mu)|_{T}=(dN/dT)|_{\mu}$. A careful measurement of the QD occupation as a function of temperature can, therefore, give experimental access to the change in entropy $\Delta S$ and thus allow for the detection of fractional entropy values \cite{PhysRevLett.123.147702}.

\section{Perturbative Renormalization Group Analysis}\label{Poor_man_scaling}
\subsection{The effective Hamiltonian}
Since we are now interested in the Kondo effect in the system which appears when the QD have a finite magnetic moment, let us look to the subspace of the total Hamiltonian that embodies only the singly occupied states of the QD. To this end, we follow the projection method \cite{Hewson}, which is equivalent to performing a Schrieffer-Wolff transformation. A Similar approach has been employed in an analogue system composed by MZMs coupled to an interacting QD~\cite{PhysRevB.94.045316}. After a somewhat cumbersome but straightforward calculation, the resulting effective Hamiltonian can be written as
\begin{eqnarray}\label{H_eff}
H_{\rm eff}\!=\!H_{\leads}+H_{\rm K}+H_{\gamma_{r}}+H_{\gamma_{l}}+H_{\gamma_{lr}}+
 H_{\lambda_{r}^{2}}+H_{\lambda_{l}^{2}},\quad
\end{eqnarray}
where the individual terms in \eqref{H_eff} are given by:
\begin{eqnarray}
H_{\rm K}&=&\sum_{\ell\veck\veck'}J_{\ell\veck\veck'}[S^{z}(c_{\ell\veck'\uparrow}^{\dag}c_{\ell\veck\uparrow}-
c_{\ell\veck'\downarrow}^{\dag}c_{\ell\veck\downarrow}) \nonumber \\
 & &+S^{+}c_{\ell\veck'\downarrow}^{\dag}c_{\ell\veck\uparrow}+S^{-}c_{\ell\veck'\uparrow}^{\dag}
c_{\ell\veck\downarrow}] \; ,
\end{eqnarray}
\begin{eqnarray}
\label{Hgr}
 H_{\gamma_{r}}&=&\sum_{\veck}(\Upsilon_{r\veck}S^{-} 
\gamma_{r} c_{\veck\downarrow}  +\Upsilon_{r\veck}^{*} S^{+}
c_{\veck \downarrow}^{\dag}\gamma_{r}) \nonumber \\
&&-\sum_{\veck}(\hat{T}_{r\veck}\gamma_{r} c_{\veck\uparrow}+ 
\hat{T}_{r\veck}^{*}c_{\veck\uparrow}^{\dag} \gamma_{r}) \; ,
\end{eqnarray}
\begin{eqnarray}
\label{Hgl}
H_{\gamma_{l}}&=&\sum_{\veck}(\Upsilon_{l\veck}S^{+}\gamma_{l}c_{
\veck\uparrow} +\Upsilon_{l\veck}^{*}S^{-}c_{ \veck 
\uparrow}^{\dag}\gamma_{l}) \nonumber \\
&&-\sum_{\veck}(\hat{T}_{l\veck}\gamma_{l} 
c_{\veck\downarrow} + \hat{T}_{l\veck}^{*}c_{\veck\downarrow}^{\dag} 
\gamma_{l}) \; ,
\end{eqnarray}
\begin{equation}\label{H_lr}
 H_{\gamma_{lr}}=\Upsilon_{lr}S^{-}\gamma_{l}\gamma_{r}+\Upsilon_{lr}^{*}S^{+}\gamma_{r}\gamma_{l} \; ,
\end{equation}
\begin{equation}
 H_{\lambda_{r}^{2}}=\lambda_{r}^{2}\left(\frac{n_{d\uparrow}}{\varepsilon_{d}}
 -\frac{n_{d\downarrow}}{\varepsilon_{d}+U}\right) \; ,
\end{equation}
and
\begin{equation}
 H_{\lambda_{l}^{2}}=\lambda_{l}^{2}\left(\frac{n_{d\downarrow}}{\varepsilon_{d}}-
 \frac{n_{d\uparrow}}{\varepsilon_{d}+U}\right) \; .
\end{equation}

In the above equations, we used the standard spin-charge relations $S_{z} = (n_{d\uparrow}-n_{d\downarrow})/2$; $S^{+}=d_{\uparrow}^{\dag}d_{\downarrow}$ and 
$S^{-}=d_{\downarrow}^{\dag}d_{\uparrow}$. The couplings are given by
\begin{equation}\label{J_kkp}
 J_{\veck\veck'}=V_{\veck}V_{\veck'}^{*}\left(\frac{1}{\varepsilon_{\veck}-\varepsilon_{d}}+\frac{1}{\varepsilon_{d}+U-\varepsilon_{\veck'}}\right) \; ,
\end{equation}
\begin{equation}
 \Upsilon_{r\veck}=-\lambda_{r}V_{\veck}\left(\frac{1}{\varepsilon_{\veck}-\varepsilon_{d}}
 +\frac{1}{\varepsilon_{d}+U-\varepsilon_{\veck}}\right)e^{i\theta} \; ,
\end{equation}
\begin{equation}
 \hat{T}_{r\veck}=-\lambda_{r}V_{\veck}\left(\frac{n_{d\uparrow}}{\varepsilon_{d}-\varepsilon_{\veck}}
 +\frac{n_{d\downarrow}}{\varepsilon_{d}+U-\varepsilon_{\veck}}\right)e^{i\theta} \; ,
\end{equation}
\begin{equation}
 \Upsilon_{l\veck}=\lambda_{l}V_{\veck}\left(\frac{1}{\varepsilon_{\veck}-\varepsilon_{d}}
 +\frac{1}{\varepsilon_{d}+U-\varepsilon_{\veck}}\right) \; ,
\end{equation}
\begin{equation}
  \hat{T}_{l\veck}=\lambda_{r}V_{\veck}\left(\frac{n_{d\downarrow}}{\varepsilon_{d}-\varepsilon_{\veck}}
 +\frac{n_{d\uparrow}}{\varepsilon_{d}+U-\varepsilon_{\veck}}\right) \; ,
\end{equation}
and
\begin{equation}
 \Upsilon_{lr}=\lambda_{r}\lambda_{l}e^{i\theta}\left(\frac{1}{U+\varepsilon_{d}}-\frac{1}{\varepsilon_{d}}\right) \; .
\end{equation}
In the above, we  have introduced $\theta \! \equiv \! \delta\phi/2+\pi/2$. Rigorously, the dependence of the above  on $\veck$ and $\veck^\prime$  in Eq.~\eqref{J_kkp} should be symmetrized. However, this is not necessary as we now assume that the effective couplings depend  weakly on $\veck$ such that $V_{\veck}=V$. We also set  $\varepsilon_{\veck}=\varepsilon_{\veck'}\approx 0$ in the couplings above.
Within these approximations, we obtain
\begin{equation}\label{J_0}
 J=|V|^{2}\left(\frac{1}{\varepsilon_{d}+U}-\frac{1}{\varepsilon_{d}}\right),
\end{equation}
which is the usual Kondo coupling and the MZM-related effective couplings given by:
\begin{equation}\label{Ur}
 \Upsilon_{r}=-\lambda_{r}V\left(\frac{1}{\varepsilon_{d}+U}-\frac{1}{\varepsilon_{d}}\right)e^{i\theta} \;,
\end{equation}
\begin{equation}\label{Ul}
 \Upsilon_{l}=\lambda_{l}V\left(\frac{1}{\varepsilon_{d}+U}-\frac{1}{\varepsilon_{d}}\right) \; ,
\end{equation}
\begin{equation}\label{Tr}
 \hat{T}_{r}=-\lambda_{r}V\left(\frac{n_{d\uparrow}}{\varepsilon_{d}}
 +\frac{n_{d\downarrow}}{\varepsilon_{d}+U}\right)e^{i\theta} \; ,
\end{equation}
\begin{equation}\label{Tl}
  \hat{T}_{l}=\lambda_{r}V\left(\frac{n_{d\downarrow}}{\varepsilon_{d}}
 +\frac{n_{d\uparrow}}{\varepsilon_{d}+U}\right) \; .
\end{equation}
Extra caution is needed in these last two effective couplings as  they are not c-numbers like the previous ones. For instance,  let us look at them  near the particle-hole symmetry point ($\e_d=-U/2$). To this end, let us rewrite the dot level as $\varepsilon_{d}=-U/2+\delta \varepsilon$, with $\delta\e \ll U$ . We can then write 
\begin{eqnarray*}
 \hat{T}_{r}\!&=&\!-2\lambda_{r}V\!\!\left[-\frac{n_{d\uparrow}}{U}\left(1-\frac{2\delta \varepsilon}{U}\right)^{-1}\!\!\!+\!
 \frac{n_{d\downarrow}}{U}\left(1+\frac{2\delta \varepsilon}{U}\right)^{-1}\right]e^{i\theta}.
\end{eqnarray*}
Using the approximation $(1-x)^{-1}\approx 1+x$, and $(1+x)^{-1}\approx 1-x$, for $x \ll  1$, we obtain 
\begin{eqnarray*}
\hat{T}_{r}\approx-\lambda_{r}V\left[-\frac{4}{U}S_{z}-\frac{4\delta \varepsilon}{U^{2}}\right]e^{i\theta}.
\end{eqnarray*}
Here we have used $S_{z}=(n_{d\uparrow}-n_{d\downarrow})/2$ and 
$n_{d\uparrow}+n_{d\downarrow}=1$. With this expression, we can write 
\begin{eqnarray}\label{H_Tr}
\!\!\!\!\! \sum_{\veck}(\hat{T}_{r}\gamma_{r}c_{\veck\uparrow}+
\hat{T}_{r}^{*}c_{\veck\uparrow}^{\dag}\gamma_{r})&=&\sum_{\veck}\left[({T}_{r}\gamma_{r}c_{\veck\uparrow}+
{T}_{r}^{*}c_{\veck\uparrow}^{\dag}\gamma_{r})\right.\nonumber\\
& &\left.-({T}_{rz}S_{z}\gamma_{r}c_{\veck\uparrow}+
{T}_{rz}^{*}S_{z}c_{\veck\uparrow}^{\dag}\gamma_{r})\right], \nonumber \\
\end{eqnarray}
where we have 
\begin{equation}
 T_{r} \approx \frac{4\delta \varepsilon}{U^{2}}\lambda_{r}Ve^{i\theta},\quad \mbox{and} \quad T_{rz}=-\frac{4}{U}\lambda_{r}Ve^{i\theta}.
\end{equation}
In a similar fashion, we can write
\begin{equation}
 \hat{T}_{l}\approx \lambda_{l}V\left[\frac{4}{U}S_{z}-\frac{4\delta \varepsilon}{U^{2}}\right],
\end{equation}
to obtain
\begin{eqnarray}\label{H_Tl}
\!\!\!\!\!  \sum_{\veck}(\hat{T}_{l}\gamma_{r}c_{\veck\downarrow}+
\hat{T}_{l}^{*}c_{\veck\downarrow}^{\dag}\gamma_{l})&=&\sum_{\veck}\left[({T}_{l}\gamma_{l}c_{\veck\downarrow}+
{T}_{l}^{*}c_{\ell\veck\downarrow}^{\dag}\gamma_{l})\right.\nonumber\\
& &\left.+({T}_{lz}S_{z}\gamma_{l}c_{\veck\downarrow}+
{T}_{lz}^{*}S_{z}c_{\veck\downarrow}^{\dag}\gamma_{l})\right],\nonumber \\
\end{eqnarray}
with
\begin{equation}
 T_{l}=-\frac{4\delta \varepsilon}{U^{2}}\lambda_{l}V,\quad \mbox{and } \quad  T_{lz}=\frac{4}{U}\lambda_{l}V.
\end{equation}
With the help of Eqs.~\eqref{H_Tr}  and \eqref{H_Tl}, we can see that the Hamiltonians  \eqref{Hgr} and \eqref{Hgl} describe a sort of ``triple exchange interaction'' involving the  spins of the the electrons in the  QD, the spins of the electrons in the conduction band and the Majorana modes. More interestingly, notice  that Eq.\ \eqref{H_lr} describes a local exchange interaction between the QD and the two Majorana modes, favoring the formation of a local singlet, thereby competing with the Kondo effect. However, as we will see below, these interactions turn out to be only marginally relevant in the RG sense  and therefore do not affect the flow to the Kondo fixed point of the Hamiltonian, at least at the perturbed renormalization group level.

\subsection{Poor-man scaling analysis}
\label{sec:numerical}
To compute the low-energy fixed point of the effective Hamiltonian, we  employ the perturbative Anderson's poor man's scaling analysis \cite{Anderson1970}. This method consists in integrating out the high-energy electronic states at the edge of conduction band, providing us with a set of effective couplings as functions of the reduced bandwidth parameter. For our system, the calculations are straightforward but quite involved. As a result, we obtain the following scaling equations
\begin{subequations}
\label{diff_eq}
 \begin{eqnarray}
  \dot \Upsilon_{r}&=&-\frac{\rho}{2} \Upsilon_{r}J-\rho T_{r}J-\frac{\rho}{2} T_{rz}J \;,  
  \label{eq:Upsilonrflow}\\ 
  \dot T_{rz}&=&-\rho\Upsilon_{r}J-\rho T_{r}J \;,\\
  \dot \Upsilon_{lr}&=&-\rho \Upsilon_{r}  T_{l}- 
\rho\Upsilon_{l} T_{r} -\frac{\rho}{2} T_{rz} \Upsilon_{l}
-\frac{\rho}{2} T_{lz} \Upsilon_{r} \;,\\
\dot T_{r}&=&-\frac{\rho}{2} \Upsilon_{r}J-\frac{\rho}{4}T_{rz}J \;,
\label{eq:Trflow}\\
  \dot \Upsilon_{l}&=&-\frac{\rho}{2} \Upsilon_{l} J-\rho T_{l}J-\frac{\rho}{2} T_{lz} J \;,
  \label{eq:Upsilonlflow}\\ 
 \dot T_{lz} &=&-\rho \Upsilon_{l} J-\rho T_{l} J \;,\\
  \dot T_{l} &=&-\frac{\rho}{2} \Upsilon_{l} J-\frac{\rho}{4} T_{lz} J\;,\\
  \dot J&=&-2\rho J^{2} \label{eq:Jflow}.
 \end{eqnarray}
\end{subequations}

In the above equations, we have defined $\dot X \equiv d X/d\ln \tilde D$, where $\tilde D$ in the reduced bandwidth. We have also denoted the density of states of the conduction electrons calculated and the Fermi level ($\e_F\!=\!0$) as $\rho \!=\! \rho(0)$.
Notice  that the differential equation for $J$ is completely decoupled from the rest of the RG flow, and is equivalent to the usual isotropic Kondo model. Therefore, this clearly indicates that, regardless the coupling of the quantum impurity to the Majorana modes, the system will always flow to the usual strongly coupling screening fixed point, where the QD is  Kondo screened by the conduction electrons band as the temperature decreases characteristic Kondo temperature. The fact that this fixed point is stable even in the presence of arbitrary MZM-QD coupling is one of the central results of the paper.

To  obtain the evolution of the other couplings with $\tilde D$, the system of coupled differential equations above has to be solved, given a initial condition for the couplings. Note that only the couplings with subindex $r$ can have complex initial values.  Since Eq.\ \eqref{eq:Jflow} is fully decoupled from the others, given a real initial condition $J(0)$, the coupling $J$  will remain real along renormalization flow. Notice also that the Eqs.\ \eqref{eq:Upsilonlflow}-\eqref{eq:Jflow} are fully decoupled from the others and thus have pure real solutions. It is now easy to see that the real (imaginary) part of the solution of Eqs.\ \eqref{eq:Upsilonrflow}-\eqref{eq:Trflow} depends only on the real (imaginary) part of their initial conditions.

As  there are several model parameters which can be tuned, we will resort to a numerical solution of the system of differential equations \eqref{diff_eq} within a few specific situations of interest and look at the evolution of the renormalized parameters as a function of the band width cut-off running parameter $\tilde D$.  As in Sec.~\ref{sec:NRG}, we set $D=1$, $U=0.5$, $\delta=0$ (the particle-hole symmetric case) and $V=0.1$. We also set $\delta\phi=\pi$, for which case the couplings are all real (different values of $\delta\phi$ lead to qualitative  equivalent results).  

\begin{figure}[t]
\centering
\subfigure{\includegraphics[clip,width=3.2in]{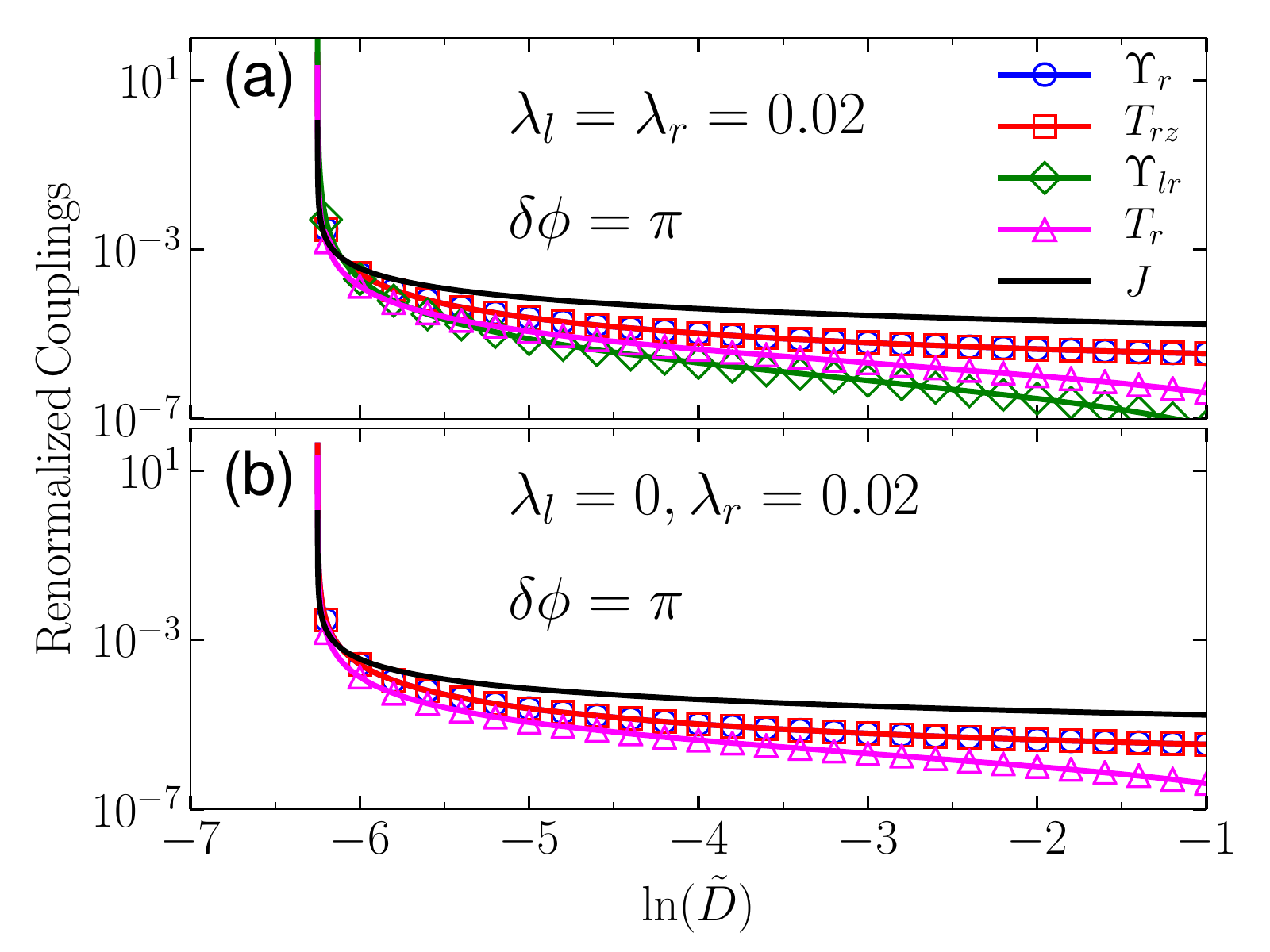}}
\caption{(Color online)~Renormalized parameters as function of bandwidth cut-off $\tilde D$ for (a) symmetric case, $\lambda_l=\lambda_r=0.02$ and (b) asymmetric case, $\lambda_r=0.02$ and $\lambda_l=0$. The other parameters are $\delta\phi=\pi$, $U=0.5$, $V=0.1$, $\delta=0$. These parameters correspond to those of diamonds (green) and circles (red) curves, respectively, of Fig.~\ref{entropy_NRG}. All couplings are normalized by the value of $J$ right before the break down of the perturbative procedure.} 
\label{couplings-1}
\end{figure}

The results for the calculation are shown in Fig.~\ref{couplings-1}(a) for $\lambda_l=\lambda_r=0.02$  (symmetric) and Fig.~\ref{couplings-1}(b) for $\lambda_l=0$ and $\lambda_r=0.02$ (asymmetric) cases.  These choices correspond to the diamond (green) and circle (red) curves of Fig.~\ref{entropy_NRG}, respectively. 

We first note that  all the parameters, including $J$, renormalize to infinity as   $\tilde D\rightarrow 0$. This indicates that the system evolves towards a strong coupling fixed point in which all the couplings of the effective model (Eq.\ \eqref{H_eff}) diverge, which is consistent with our NRG calculations of Sec.~\ref{sec:NRG}.

One can now ask how the couplings of the QD with the MZMs affects the Kondo temperature of the system. By extracting $T_K$ from the value of $\tilde D$ for which $J\rightarrow \infty$, we can solve the Eq.~\eqref{eq:Jflow} separately, obtaining exactly the same Kondo temperature for a single impurity Kondo model, $k_BT_{\rm K}=D\,\exp{(-1/\rho_0J_0)}$, with $J_0$ given by  Eq.~\eqref{J_0}. Additionally, we could  also chose to numerically extract $T_{\rm K}$ from the value of $\tilde D$ for which the renormalization flow  of entire system of equations \eqref{diff_eq} breaks down. By performing this procedure, we see no clear dependence of $T_{\rm K}$ with $\lambda$. We can see, however, that \textit{all} couplings diverge at the same scale for which  $J$ diverges (see  Fig.~\ref{couplings-1}). Since $\Upsilon\rightarrow \infty$ at the fixed point, this indicates that there is indeed a type of local strong interaction between the spins of the dot and the Majorana operators. Nevertheless, such effective coupling does not affect the Kondo screening.

More importantly, the fact that the renormalized couplings, $\Upsilon_{r(l)}$, $\Upsilon_{lr}$, $T_{r(l)}$ and $T_{r(l)z}$ also diverge in the Kondo-screened fixed point ($J\rightarrow \infty$) suggests some sort of magnetic interaction between the Majorana modes and the QD.  This interpretation relies solely on the form of the Hamiltonian terms $H_{\gamma_r}$, $H_{\gamma_l}$ and $H_{\gamma_lr}$ of the effective model $H_{\rm eff}$~\eqref{H_eff}. From the RG perspective, the divergence of the renormalized couplings is associated with a marginally relevant local interaction between the Majorana modes and the QD 
spin. 

\section{Concluding remarks}
\label{sec:conclusions}

To summarize, we studied a system composed of a quantum dot in the Kondo regime coupled to two Majorana zero modes with opposite spin polarizations. Our numerical renormalization group calculations show that the low-energy features are determined by the high-energy MZM-QD couplings in a surprisingly simple way: by counting the number of uncoupled MZMs $N_0$, one can infer the low-temperature residual entropy of the QD+MZM system as $S_{\rm res} \rightarrow (N_0/2) k_B\ln(2)$. This is true even in the noninteracting regime, which reinforces that the Kondo physics in the interacting regime remains independently of the MZM-QD couplings.
We surmise that this result can be generalized for a system with $N \leq N_0$ MZMs coupled to the dot. As discussed in the main text, we believe that this result can be experimentally checked using, e.g.,  the MZM entropy measurement protocols proposed in Ref.\ \cite{PhysRevLett.123.147702}. 

One interesting consequence is the non-Fermi liquid behavior for odd values of $N_0$. This result indicates that the strongly coupled fixed point describes a combination of a Kondo-like singlet and the free MZM modes. This picture is reinforced by a perturbative RG analysis of the effective Kondo-like low-energy Hamiltonian. The poor-man scaling analysis shows that the scaling equation for the Kondo coupling $J$ is decoupled from the others, reinforcing the fact that the stable, low-energy fixed point is Kondo-like. Nonetheless, the effective MZM-related couplings do depend on $J$, suggesting a MZM-mediated spin exchange coupling contribution. 

We believe these results are  particularly significant for future theoretical and  experimental investigations of MZMs in quantum dots coupled to topological superconductors, as it sheds light on the intricate interplay between MZMs and Kondo effect in these systems.

\acknowledgements
We thank Rok Zitko for valuable discussions. This work was supported by the Brazilian funding agencies FAPEMIG, CAPES,  CNPq (Grant Nos.  423137/2018-2, 305738/2018-6), and FAPESP (Grant No. 2016/18495-4). EV thanks support from Ohio University within the Robert Glidden Visiting Professorship program.

\appendix
\section{Analytical results for the residual entropy in the non-interacting regime}
\label{sec:analytic}
\subsection{Free Majorana mode entropy}
\noindent

Let us consider a system composed of a resonant level with a regular fermion~(spinless for simplicity), which can be described  by two Majorana modes, coupled to a metallic lead. Here, only one of them is coupled to the lead, while the another one remains free. The Hamiltonian of this toy model is given by 
\begin{equation}\label{H}
 H=\sum_{\mathbf{k}}\varepsilon_{k}c_{\mathbf{k}}^{\dag}c_{\mathbf{k}}+\sum_{\mathbf{k}}
 (Vc_{\mathbf{k}}\gamma_{1}+V^{*}\gamma_{1}c_{\mathbf{k}}^{\dag}),
\end{equation}
where $V$ is the hybridization matrix element of the coupled of the Majorana mode with the metallic lead. We introduce the Green's  function of the level
\begin{equation}
 G_{f}(\omega)\equiv \langle\langle f;f^{\dag}\rangle\rangle_{\omega},
\end{equation}
where, the fermion operators can always be written as $f=(\gamma_{1}+i\gamma_{2})/\sqrt{2}$. To distinguish the contribution of the Majorana modes to the entropy is convenient to rewrite the Green's function in terms of the Majorana operators which results 
\begin{equation}\label{gd}
G_{f}(\omega)=\frac{1}{2}\left[M_{11}(\omega)-iM_{12}(\omega)+iM_{21}(\omega) +M_{22}(\omega)\right].
\end{equation}
Here we have define the Majorana Green functions 
\begin{equation}
M_{ij}(\omega)\equiv \langle\langle \gamma_{i};\gamma_{j}\rangle\rangle_{\omega}.
\end{equation}
The free energy of the level can be written as
\begin{eqnarray}\label{free}
F_{f}(T)=-k_{\rm B}T\int_{-\infty}^{\infty}\rho_{f}(\omega){\rm ln}(1+e^{-\beta\omega})d\omega
\end{eqnarray}
in which $\beta=1/k_{\rm B}T$ with $k_B$ being the Boltzmann constant and $T$ the temperature  and
\begin{equation}\label{rho}
 \rho_{f}(\omega)=-\frac{1}{\pi} {\rm Im}\{ G_{f}(\omega)\}.
\end{equation}

We can now compute the entropy using the relation 
\begin{equation}\label{entropy}
 S_{f}(T)=-\frac{\partial F_{f}(T)}{\partial T}.
\end{equation}

analytical expression for the the Majorana Green functions can be calculated using the equation of motion~(EOM)\cite{Sov.Phys.Usp..3.320}
\begin{equation}\label{motion}
\omega\langle\langle \gamma_{i};\gamma_{j}\rangle\rangle_{\omega}=\delta_{i,j}+
\langle\langle [\gamma_{i},H]_{-};\gamma_{j}\rangle\rangle_{\omega}.
\end{equation}

There are two distinct regimes for the toy model \eqref{H}, $V=0$ and $V\neq 0$, let us take a closer look in them. 

\subsubsection{Decoupled regime~(V=0)}
\noindent

In the decoupled regime the Hamiltonian \eqref{H} describes a free regular fermion, so using the EOM~\eqref{motion} we obtain  
\begin{eqnarray}
 G_{f}(\omega)=\frac{1}{2}\left[\frac{1}{\omega+i\eta}+\frac{1}{\omega+i\eta}\right],
\end{eqnarray}
where $\eta\rightarrow 0^{+}$, which results from \eqref{rho} in the density of states 
\begin{equation}\label{rd}
 \rho_{f}(\omega)=\frac{1}{2}\left[\delta(\omega)+\delta(\omega)\right],
\end{equation}
where the first term is from the contribution of the Majorana mode $\gamma_{1}$ and the second is  from $\gamma_{2}$. Notice that in this case both Majorana modes are free. Inserting \eqref{rd} into Eqs.~\eqref{free} and using \eqref{entropy} we find
\begin{eqnarray}
 F_{f}(T)=-k_{\rm B}T{\rm ln(2)}
\end{eqnarray}
and
\begin{equation}
 S_{f}(T)=k_{\rm B}{\rm ln(2)}. 
\end{equation}

This is the entropy expected for the single zero-energy level for spinless fermion.

\subsubsection{Coupled regime~($V\neq 0$)}
\noindent

In the coupled regime, using Eqs.~\eqref{H}, \eqref{gd} and \eqref{motion} we obtain the following Green function 
\begin{equation}\label{gd2}
 G_{f}(\omega)=\frac{1}{2}\left(\frac{1}{\omega+i\Gamma}+\frac{1}{\omega+i\eta}\right).
\end{equation}
Here, $\Gamma=2\pi\rho_{0}|V|^{2}$, $\rho_{0}$ is the density of states of the metallic lead. Equation \eqref{gd2} render the density of states 
\begin{eqnarray}
 \rho_{f}(\omega)
 =\frac{1}{2}\left[\frac{1}{\pi}\frac{\Gamma}{\omega^{2}+\Gamma^{2}}+\delta(\omega)\right].
\end{eqnarray}
This density of state is similar to that derived in Ref. \cite{PhysRevLett.123.147702}. The Lorentzian form of the fist term is due the Majorana coupled to the metallic lead, whereas the $\delta$-function in second term results from the free Majorana mode. Now free energy acquires the form
\begin{eqnarray}\label{free2}
 F_{f}(T)&=& F_{1}(T)+F_{2}(T) \nonumber \\
 &=&-\frac{k_{\rm B}T}{2}\int_{-\infty}^{\infty}
 \frac{1}{\pi}\frac{\Gamma}{\omega^{2}+\Gamma^{2}}{\rm ln}(1+e^{-\beta\omega})d\omega \nonumber \\
 &&-\frac{k_{\rm B}T}{2}\rm ln(2).
\end{eqnarray}
With this, using Eq.~\eqref{entropy} we find
\begin{eqnarray}\label{entropy2}
 S_{f}(T)&=&S_{1}(T)+S_{2}(T) \nonumber \\
 &=&\frac{k_{\rm B}}{2}\int_{-\infty}^{\infty}
 \frac{1}{\pi}\frac{\Gamma}{\omega^{2}+\Gamma^{2}}\frac{\partial}{\partial T} \left[T{\rm ln}(1+e^{-\beta\omega})\right]d\omega \nonumber \\ 
 &&+\frac{k_{\rm B}}{2}\rm ln(2). 
 \end{eqnarray}

In the above, $S_2(T)=(1/2)k_{\rm B} \rm ln(2)$ corresponds to the contribution from the free Majorana mode, $\gamma_2$. The contribution $S_{1}(T)$ (given by the second line of the equation above) from the Majorana mode, $\gamma_1$,  coupled to the metallic lead requires more attention. Performing the temperature derivative we  obtain
\begin{eqnarray}\label{S1}
 S_{1}(T)&=&\frac{k_{\rm B}}{2}\int_{-\infty}^{\infty}
 \frac{1}{\pi}\frac{\Gamma}{\omega^{2}+\Gamma^{2}}\frac{\partial}{\partial T}
 \left[T{\rm ln}(1+e^{-\beta\omega})\right]d\omega  \nonumber \\
 &=&\frac{k_{\rm B}}{2}\int_{-\infty}^{\infty}
 \frac{1}{\pi}\frac{\Gamma}{\omega^{2}+\Gamma^{2}}{\rm ln}(1+e^{-\beta\omega})d\omega \nonumber \\
 &&\quad+\frac{1}{2T}\int_{-\infty}^{\infty}
 \frac{1}{\pi}\frac{\Gamma}{\omega^{2}+\Gamma^{2}}\frac{\omega}{(e^{\beta\omega}+1)}d\omega. 
\end{eqnarray}

The integrals of Eq.~\eqref{S1} are rather complicated. Yet, approximated analytical expressions can be obtained in the low-temperature limit. First, note that since ${\rm ln}(1+e^{-\beta\omega})\rightarrow 0$ for and $\Gamma \gg k_{\rm B}T$ as $T\rightarrow 0$. With this we can see that 
$$\int_{-\infty}^{\infty}
 \frac{1}{\pi}\frac{\Gamma}{\omega^{2}+\Gamma^{2}}{\rm ln}(1+e^{-\beta\omega})d\omega \rightarrow 0
 ,\qquad (T\rightarrow 0).$$
Moreover, using the fact that $f(\omega)=(e^{\beta\omega}+1)^{-1}$ we can write
\begin{equation}
 S_{1}(T\rightarrow 0)=\int_{-\infty}^{\infty}h(\omega)f(\omega)d\omega,
\end{equation}
with 
\begin{eqnarray}\label{h}
h(\omega)=\frac{1}{2\pi T}\frac{\omega\Gamma}{\omega^{2}+\Gamma^{2}}
\end{eqnarray}

For small $T$ this class of integrals can be written in within the so called Sommerfeld expansion as
\begin{equation}\label{som}
 S_{1}(T)\approx g(0)+\frac{\pi^{2}}{6}\frac{d^{2}g(\omega)}{d\omega^{2}}\Big|_{\omega=0}
 (k_{\rm B}T)^{2}+\mathcal{O}[(k_{\rm B}T)^{4}],
\end{equation}
in which
\begin{equation}\label{g}
 g(\omega)=\int_{0}^{\omega}h(\omega')d\omega'.
\end{equation}
Upon performing the integration with the integrand \eqref{h} we obtain
\begin{equation}
 g(\omega)=
\frac{\Gamma}{4\pi T}{\rm ln}\left[\frac{\omega^{2}+\Gamma^{2}}{\Gamma^{2}}\right].
\end{equation}

Using this expression in \eqref{som}, up to $\mathcal{O}[(k_{\rm B}T)^{3}]$  we find
\begin{equation}
S_{1}(T)\approx\frac{k_{\rm B}}{2}\frac{\pi}{6}\left(\frac{k_{\rm B}T}{\Gamma}\right).
\end{equation}

The total entropy in the coupled case is then given by 
\begin{equation}
 S_{f}(T)\approx\frac{k_{\rm B}}{2}\rm ln(2)+\frac{k_{\rm B}}{2}\frac{\pi}{6}\left(\frac{k_{\rm B}T}{\Gamma}\right) 
,\qquad (T\rightarrow 0).
\end{equation}
Strictly at $T=0$ we obtain the interesting result
\begin{equation}
 S_{\rm res}=\frac{k_{\rm B}}{2}\rm ln(2),
\end{equation}
where $S_{\rm res}\equiv S_{f}(0)$.
This result shows that only the free Majorana mode contributes to the residual entropy at $T=0$. Furthermore the residual entropy is half of $k_B{\ln(2)}$, revealing  the non-Fermi liquid~(NFL) characteristic of the system. This is the analogue of what happens in the two channel Kondo effect.
\cite{Tsvelick_1985,Lud1,PhysRevLett.67.161,PhysRevB.48.7297,PhysRevB.46.10812,PhysRevLett.70.686,PhysRevB.50.17732,PhysRevB.52.6611,Cox}
 
\section{Connection with the effective model}\label{appendix_B}
\noindent 

In the non-interacting limit~($U=0$) we have, from \eqref{H_eff}, \eqref{J_0}-\eqref{Tl}, 
$J=\Upsilon_{lr}=\Upsilon_{r}=\Upsilon_{l}=0$ and the 
effective Hamiltonian becomes 
\begin{eqnarray}\label{H_eff2}
 H_{\rm eff}&=&\sum_{\ell\veck\sigma}\varepsilon_{\ell\veck\sigma}c_{\ell\veck\sigma}^{\dag}c_{\ell\veck\sigma}
 +\sum_{ 
\ell\veck}({T}_{r}\gamma_{r} c_{\ell\veck\uparrow}+ 
{T}_{r}^{*}c_{\ell\veck\uparrow}^{\dag} \gamma_{r}) \nonumber \\
&&\quad +\sum_{ \ell\veck}({T}_{l}\gamma_{l} 
c_{\ell\veck\downarrow} + {T}_{l}^{*}c_{\ell\veck\downarrow}^{\dag} 
\gamma_{l}),
\end{eqnarray}
with 
\begin{equation}
 T_{r}=2\lambda_{r}\left(\frac{V}{\e_{d}}\right)e^{i\theta},
 \quad T_{r}=-2\lambda_{l}\left(\frac{V}{\e_{d}}\right). 
\end{equation}
Notice that the effective Hamiltonian \eqref{H_eff2} has the same form of the Hamiltonian \eqref{H}. The main difference is that now we have  an additional degree of freedom associated with the spin. As we have seen before, the connection between the regular fermions and the Majorana Fermions can be describe as 
\begin{equation}\label{f}
 f_{\uparrow}=\frac{1}{\sqrt{2}}(\gamma_{1r}+i\gamma_{2r}),\qquad 
 f_{\downarrow}=\frac{1}{\sqrt{2}}(\gamma_{1l}+i\gamma_{2l}).
\end{equation}
The number of free MZMs is determined by  
the couplings $T_{r}$ and $T_{l}$. Let us look to the each possible configurations and their respective residual entropy.

\subsubsection{$T_{r}=T_{l}=0$}
When $T_{r}=T_{l}=0$, from Eqs.~\eqref{H_eff2} and \eqref{f} we see four free Majorana modes and, as showed in the appendix \ref{sec:analytic} each free Majorana mode contributes for the density of state with $\delta(\omega)/2$. As a result,
\begin{equation}
 \rho_{f}(\omega)=2\delta(\omega),
\end{equation}
rendering the entropy
\begin{equation}
 S_{f}(T)=2k_{\rm B}{\rm ln(2)}=S_{\rm res}.
\end{equation}

\subsubsection{$T_{r}\neq 0$ or $T_{l}\neq 0$}

In this situation we have three free Majorana modes and one coupled to the 
metallic lead which results in the density of states 
\begin{equation}
 \rho_{f}(\omega)=\frac{3}{2}\delta(\omega)+\frac{1}{2\pi}\frac{\Gamma}{\omega^{2}+\Gamma^{2}}.
\end{equation}
Using the results form Sec.~\ref{sec:analytic} we obtain the following entropy 
\begin{equation}
S_{f}(T)\approx\frac{3k_{\rm B}}{2}\rm ln(2)+
\frac{k_{\rm B}}{2}\frac{\pi}{6}\left(\frac{k_{\rm B}T}{\Gamma}\right). 
\end{equation}
At zero temperature the residual entropy
\begin{equation}
 S_{\rm res}=\frac{3k_{\rm B}}{2}\rm ln(2),
\end{equation}
from which the NFL behavior of the system is explicit.

\subsubsection{$T_{r}\neq 0$ and $T_{l}\neq 0$}

This is the  last possible situation. We now we have two free Majorana modes and two coupled Majorana modes, which provides us with the density of state 
\begin{equation}
 \rho_{f}=\delta(\omega)+\frac{1}{\pi}\frac{\Gamma}{\omega^{2}+\Gamma^{2}}.
\end{equation}
and entropy
\begin{equation}
 S_{f}(T)\approx k_{\rm B}\rm ln(2)+k_{\rm B}\frac{\pi}{6}\left(\frac{k_{\rm B}T}{\Gamma}\right). 
\end{equation}
For $T=0$ we then obtain
\begin{equation}
 S_{\rm res}=k_{\rm B}\rm ln(2).
\end{equation}
In summary, the residual entropy in the three cases can  be written as
\begin{equation}\label{entropy3}
  S_{\rm res} =
       \begin{cases}

             2k_{\rm B}{\rm ln(2)}, & \mbox{if } T_{r}=V_{l}=0. \\

             \frac{3}{2}k_{\rm B}{\rm ln(2)}, & \mbox{if } T_{r}\neq 0 \quad \mbox{or } \quad T_{r}\neq 0. \\
             k_{\rm B}{\rm ln(2)}, & \mbox{if }  T_{r}\neq 0 \quad  \mbox{and } \quad T_{l}\neq 0.
       \end{cases}
\end{equation}
Observe that the results of \eqref{entropy3} corroborate with the results obtained by NRG showed in Fig.~\ref{entropy_NRG_U}. Moreover, in  general the residual entropy can be associated with the number of free Majoranas modes~($N_{0}$) of the system as
\begin{equation}
 S_{\rm res}=\frac{N_{0}}{2}k_{\rm B}{\rm ln (2)}. 
\end{equation}


\begin{thebibliography}{53}%
	\makeatletter
	\providecommand \@ifxundefined [1]{%
		\@ifx{#1\undefined}
	}%
	\providecommand \@ifnum [1]{%
		\ifnum #1\expandafter \@firstoftwo
		\else \expandafter \@secondoftwo
		\fi
	}%
	\providecommand \@ifx [1]{%
		\ifx #1\expandafter \@firstoftwo
		\else \expandafter \@secondoftwo
		\fi
	}%
	\providecommand \natexlab [1]{#1}%
	\providecommand \enquote  [1]{``#1''}%
	\providecommand \bibnamefont  [1]{#1}%
	\providecommand \bibfnamefont [1]{#1}%
	\providecommand \citenamefont [1]{#1}%
	\providecommand \href@noop [0]{\@secondoftwo}%
	\providecommand \href [0]{\begingroup \@sanitize@url \@href}%
	\providecommand \@href[1]{\@@startlink{#1}\@@href}%
	\providecommand \@@href[1]{\endgroup#1\@@endlink}%
	\providecommand \@sanitize@url [0]{\catcode `\\12\catcode `\$12\catcode
		`\&12\catcode `\#12\catcode `\^12\catcode `\_12\catcode `\%12\relax}%
	\providecommand \@@startlink[1]{}%
	\providecommand \@@endlink[0]{}%
	\providecommand \url  [0]{\begingroup\@sanitize@url \@url }%
	\providecommand \@url [1]{\endgroup\@href {#1}{\urlprefix }}%
	\providecommand \urlprefix  [0]{URL }%
	\providecommand \Eprint [0]{\href }%
	\providecommand \doibase [0]{http://dx.doi.org/}%
	\providecommand \selectlanguage [0]{\@gobble}%
	\providecommand \bibinfo  [0]{\@secondoftwo}%
	\providecommand \bibfield  [0]{\@secondoftwo}%
	\providecommand \translation [1]{[#1]}%
	\providecommand \BibitemOpen [0]{}%
	\providecommand \bibitemStop [0]{}%
	\providecommand \bibitemNoStop [0]{.\EOS\space}%
	\providecommand \EOS [0]{\spacefactor3000\relax}%
	\providecommand \BibitemShut  [1]{\csname bibitem#1\endcsname}%
	\let\auto@bib@innerbib\@empty
	\bibitem [{\citenamefont {Kitaev}(2001)}]{Kitaev}%
	\BibitemOpen
	\bibfield  {author} {\bibinfo {author} {\bibfnamefont {A.~Y.}\ \bibnamefont
			{Kitaev}},\ }\href {\doibase 10.1070/1063-7869/44/10S/S29} {\bibfield
		{journal} {\bibinfo  {journal} {Physics-Uspekhi}\ }\textbf {\bibinfo {volume}
			{44}},\ \bibinfo {pages} {131} (\bibinfo {year} {2001})}\BibitemShut
	{NoStop}%
	\bibitem [{\citenamefont {Liu}\ \emph {et~al.}(2015)\citenamefont {Liu},
		\citenamefont {Cheng},\ and\ \citenamefont {Lutchyn}}]{Liu}%
	\BibitemOpen
	\bibfield  {author} {\bibinfo {author} {\bibfnamefont {D.~E.}\ \bibnamefont
			{Liu}}, \bibinfo {author} {\bibfnamefont {M.}~\bibnamefont {Cheng}}, \ and\
		\bibinfo {author} {\bibfnamefont {R.~M.}\ \bibnamefont {Lutchyn}},\ }\href
	{\doibase 10.1103/PhysRevB.91.081405} {\bibfield  {journal} {\bibinfo
			{journal} {Phys. Rev. B}\ }\textbf {\bibinfo {volume} {91}},\ \bibinfo
		{pages} {081405} (\bibinfo {year} {2015})}\BibitemShut {NoStop}%
	\bibitem [{\citenamefont {Alicea}(2012)}]{Alicea-1}%
	\BibitemOpen
	\bibfield  {author} {\bibinfo {author} {\bibfnamefont {J.}~\bibnamefont
			{Alicea}},\ }\href {\doibase 10.1088/0034-4885/75/7/076501} {\bibfield
		{journal} {\bibinfo  {journal} {Reports on progress in physics. Physical
				Society (Great Britain)}\ }\textbf {\bibinfo {volume} {75}},\ \bibinfo
		{pages} {076501} (\bibinfo {year} {2012})}\BibitemShut {NoStop}%
	\bibitem [{\citenamefont {Read}\ and\ \citenamefont {Green}(2000)}]{Read}%
	\BibitemOpen
	\bibfield  {author} {\bibinfo {author} {\bibfnamefont {N.}~\bibnamefont
			{Read}}\ and\ \bibinfo {author} {\bibfnamefont {D.}~\bibnamefont {Green}},\
	}\href {\doibase 10.1103/PhysRevB.61.10267} {\bibfield  {journal} {\bibinfo
			{journal} {Physical Review B}\ }\textbf {\bibinfo {volume} {61}},\ \bibinfo
		{pages} {10267} (\bibinfo {year} {2000})}\BibitemShut {NoStop}%
	\bibitem [{\citenamefont {Oreg}\ \emph {et~al.}(2010)\citenamefont {Oreg},
		\citenamefont {Refael},\ and\ \citenamefont {von Oppen}}]{Oreg}%
	\BibitemOpen
	\bibfield  {author} {\bibinfo {author} {\bibfnamefont {Y.}~\bibnamefont
			{Oreg}}, \bibinfo {author} {\bibfnamefont {G.}~\bibnamefont {Refael}}, \ and\
		\bibinfo {author} {\bibfnamefont {F.}~\bibnamefont {von Oppen}},\ }\href
	{\doibase 10.1103/PhysRevLett.105.177002} {\bibfield  {journal} {\bibinfo
			{journal} {Physical Review Letters}\ }\textbf {\bibinfo {volume} {105}},\
		\bibinfo {pages} {177002} (\bibinfo {year} {2010})}\BibitemShut {NoStop}%
	\bibitem [{\citenamefont {Nayak}\ \emph {et~al.}(2008)\citenamefont {Nayak},
		\citenamefont {Simon}, \citenamefont {Stern}, \citenamefont {Freedman},\ and\
		\citenamefont {Das~Sarma}}]{Nayak}%
	\BibitemOpen
	\bibfield  {author} {\bibinfo {author} {\bibfnamefont {C.}~\bibnamefont
			{Nayak}}, \bibinfo {author} {\bibfnamefont {S.~H.}\ \bibnamefont {Simon}},
		\bibinfo {author} {\bibfnamefont {A.}~\bibnamefont {Stern}}, \bibinfo
		{author} {\bibfnamefont {M.}~\bibnamefont {Freedman}}, \ and\ \bibinfo
		{author} {\bibfnamefont {S.}~\bibnamefont {Das~Sarma}},\ }\href {\doibase
		10.1103/RevModPhys.80.1083} {\bibfield  {journal} {\bibinfo  {journal} {Rev.
				Mod. Phys.}\ }\textbf {\bibinfo {volume} {80}},\ \bibinfo {pages} {1083}
		(\bibinfo {year} {2008})}\BibitemShut {NoStop}%
	\bibitem [{\citenamefont {Mourik}\ \emph {et~al.}(2012)\citenamefont {Mourik},
		\citenamefont {Zuo}, \citenamefont {Frolov}, \citenamefont {Plissard},
		\citenamefont {Bakkers},\ and\ \citenamefont {Kouwenhoven}}]{Mourik}%
	\BibitemOpen
	\bibfield  {author} {\bibinfo {author} {\bibfnamefont {V.}~\bibnamefont
			{Mourik}}, \bibinfo {author} {\bibfnamefont {K.}~\bibnamefont {Zuo}},
		\bibinfo {author} {\bibfnamefont {S.~M.}\ \bibnamefont {Frolov}}, \bibinfo
		{author} {\bibfnamefont {S.~R.}\ \bibnamefont {Plissard}}, \bibinfo {author}
		{\bibfnamefont {E.~P. a.~M.}\ \bibnamefont {Bakkers}}, \ and\ \bibinfo
		{author} {\bibfnamefont {L.~P.}\ \bibnamefont {Kouwenhoven}},\ }\href
	{\doibase 10.1126/science.1222360} {\bibfield  {journal} {\bibinfo  {journal}
			{Science (New York, N.Y.)}\ }\textbf {\bibinfo {volume} {336}},\ \bibinfo
		{pages} {1003} (\bibinfo {year} {2012})}\BibitemShut {NoStop}%
	\bibitem [{\citenamefont {Deng}\ \emph {et~al.}(2012)\citenamefont {Deng},
		\citenamefont {Yu}, \citenamefont {Huang}, \citenamefont {Larsson},
		\citenamefont {Caroff},\ and\ \citenamefont {Xu}}]{Deng}%
	\BibitemOpen
	\bibfield  {author} {\bibinfo {author} {\bibfnamefont {M.~T.}\ \bibnamefont
			{Deng}}, \bibinfo {author} {\bibfnamefont {C.~L.}\ \bibnamefont {Yu}},
		\bibinfo {author} {\bibfnamefont {G.~Y.}\ \bibnamefont {Huang}}, \bibinfo
		{author} {\bibfnamefont {M.}~\bibnamefont {Larsson}}, \bibinfo {author}
		{\bibfnamefont {P.}~\bibnamefont {Caroff}}, \ and\ \bibinfo {author}
		{\bibfnamefont {H.~Q.}\ \bibnamefont {Xu}},\ }\href {\doibase
		10.1021/nl303758w} {\bibfield  {journal} {\bibinfo  {journal} {Nano Letters}\
		}\textbf {\bibinfo {volume} {12}},\ \bibinfo {pages} {6414} (\bibinfo {year}
		{2012})}\BibitemShut {NoStop}%
	\bibitem [{\citenamefont {Das}\ \emph {et~al.}(2012)\citenamefont {Das},
		\citenamefont {Most}, \citenamefont {Oreg}, \citenamefont {Heiblum},\ and\
		\citenamefont {Shtrikman}}]{Das}%
	\BibitemOpen
	\bibfield  {author} {\bibinfo {author} {\bibfnamefont {A.}~\bibnamefont
			{Das}}, \bibinfo {author} {\bibfnamefont {Y.}~\bibnamefont {Most}}, \bibinfo
		{author} {\bibfnamefont {Y.}~\bibnamefont {Oreg}}, \bibinfo {author}
		{\bibfnamefont {M.}~\bibnamefont {Heiblum}}, \ and\ \bibinfo {author}
		{\bibfnamefont {H.}~\bibnamefont {Shtrikman}},\ }\href {\doibase
		10.1038/nphys2479} {\bibfield  {journal} {\bibinfo  {journal} {Nature
				Physics}\ }\textbf {\bibinfo {volume} {8}},\ \bibinfo {pages} {887} (\bibinfo
		{year} {2012})}\BibitemShut {NoStop}%
	\bibitem [{\citenamefont {Sau}\ \emph {et~al.}(2010)\citenamefont {Sau},
		\citenamefont {Lutchyn}, \citenamefont {Tewari},\ and\ \citenamefont
		{Das~Sarma}}]{Sau2}%
	\BibitemOpen
	\bibfield  {author} {\bibinfo {author} {\bibfnamefont {J.~D.}\ \bibnamefont
			{Sau}}, \bibinfo {author} {\bibfnamefont {R.~M.}\ \bibnamefont {Lutchyn}},
		\bibinfo {author} {\bibfnamefont {S.}~\bibnamefont {Tewari}}, \ and\ \bibinfo
		{author} {\bibfnamefont {S.}~\bibnamefont {Das~Sarma}},\ }\href {\doibase
		10.1103/PhysRevLett.104.040502} {\bibfield  {journal} {\bibinfo  {journal}
			{Phys. Rev. Lett.}\ }\textbf {\bibinfo {volume} {104}},\ \bibinfo {pages}
		{040502} (\bibinfo {year} {2010})}\BibitemShut {NoStop}%
	\bibitem [{\citenamefont {Alicea}(2010)}]{Alicea2}%
	\BibitemOpen
	\bibfield  {author} {\bibinfo {author} {\bibfnamefont {J.}~\bibnamefont
			{Alicea}},\ }\href {\doibase 10.1103/PhysRevB.81.125318} {\bibfield
		{journal} {\bibinfo  {journal} {Phys. Rev. B}\ }\textbf {\bibinfo {volume}
			{81}},\ \bibinfo {pages} {125318} (\bibinfo {year} {2010})}\BibitemShut
	{NoStop}%
	\bibitem [{\citenamefont {Liu}\ and\ \citenamefont
		{Baranger}(2011)}]{PhysRevB.84.201308}%
	\BibitemOpen
	\bibfield  {author} {\bibinfo {author} {\bibfnamefont {D.~E.}\ \bibnamefont
			{Liu}}\ and\ \bibinfo {author} {\bibfnamefont {H.~U.}\ \bibnamefont
			{Baranger}},\ }\href {\doibase 10.1103/PhysRevB.84.201308} {\bibfield
		{journal} {\bibinfo  {journal} {Phys. Rev. B}\ }\textbf {\bibinfo {volume}
			{84}},\ \bibinfo {pages} {201308} (\bibinfo {year} {2011})}\BibitemShut
	{NoStop}%
	\bibitem [{\citenamefont {Leijnse}(2014)}]{Leijnse_2014}%
	\BibitemOpen
	\bibfield  {author} {\bibinfo {author} {\bibfnamefont {M.}~\bibnamefont
			{Leijnse}},\ }\href {\doibase 10.1088/1367-2630/16/1/015029} {\bibfield
		{journal} {\bibinfo  {journal} {New Journal of Physics}\ }\textbf {\bibinfo
			{volume} {16}},\ \bibinfo {pages} {015029} (\bibinfo {year}
		{2014})}\BibitemShut {NoStop}%
	\bibitem [{\citenamefont {Ueda}\ and\ \citenamefont
		{Yokoyama}(2014)}]{UEDA2014182}%
	\BibitemOpen
	\bibfield  {author} {\bibinfo {author} {\bibfnamefont {A.}~\bibnamefont
			{Ueda}}\ and\ \bibinfo {author} {\bibfnamefont {T.}~\bibnamefont
			{Yokoyama}},\ }\href {\doibase https://doi.org/10.1016/j.phpro.2014.09.045}
	{\bibfield  {journal} {\bibinfo  {journal} {Physics Procedia}\ }\textbf
		{\bibinfo {volume} {58}},\ \bibinfo {pages} {182 } (\bibinfo {year}
		{2014})},\ \bibinfo {note} {proceedings of the 26th International Symposium
		on Superconductivity (ISS 2013)}\BibitemShut {NoStop}%
	\bibitem [{\citenamefont {Ricco}\ \emph {et~al.}(2018)\citenamefont {Ricco},
		\citenamefont {Campo}, \citenamefont {Shelykh},\ and\ \citenamefont
		{Seridonio}}]{PhysRevB.98.075142}%
	\BibitemOpen
	\bibfield  {author} {\bibinfo {author} {\bibfnamefont {L.~S.}\ \bibnamefont
			{Ricco}}, \bibinfo {author} {\bibfnamefont {V.~L.}\ \bibnamefont {Campo}},
		\bibinfo {author} {\bibfnamefont {I.~A.}\ \bibnamefont {Shelykh}}, \ and\
		\bibinfo {author} {\bibfnamefont {A.~C.}\ \bibnamefont {Seridonio}},\ }\href
	{\doibase 10.1103/PhysRevB.98.075142} {\bibfield  {journal} {\bibinfo
			{journal} {Phys. Rev. B}\ }\textbf {\bibinfo {volume} {98}},\ \bibinfo
		{pages} {075142} (\bibinfo {year} {2018})}\BibitemShut {NoStop}%
	\bibitem [{\citenamefont {Campo}\ \emph {et~al.}(2017)\citenamefont {Campo},
		\citenamefont {Ricco},\ and\ \citenamefont {Seridonio}}]{PhysRevB.96.045135}%
	\BibitemOpen
	\bibfield  {author} {\bibinfo {author} {\bibfnamefont {V.~L.}\ \bibnamefont
			{Campo}}, \bibinfo {author} {\bibfnamefont {L.~S.}\ \bibnamefont {Ricco}}, \
		and\ \bibinfo {author} {\bibfnamefont {A.~C.}\ \bibnamefont {Seridonio}},\
	}\href {\doibase 10.1103/PhysRevB.96.045135} {\bibfield  {journal} {\bibinfo
			{journal} {Phys. Rev. B}\ }\textbf {\bibinfo {volume} {96}},\ \bibinfo
		{pages} {045135} (\bibinfo {year} {2017})}\BibitemShut {NoStop}%
	\bibitem [{\citenamefont {Gharavi}\ \emph {et~al.}(2016)\citenamefont
		{Gharavi}, \citenamefont {Hoving},\ and\ \citenamefont
		{Baugh}}]{PhysRevB.94.155417}%
	\BibitemOpen
	\bibfield  {author} {\bibinfo {author} {\bibfnamefont {K.}~\bibnamefont
			{Gharavi}}, \bibinfo {author} {\bibfnamefont {D.}~\bibnamefont {Hoving}}, \
		and\ \bibinfo {author} {\bibfnamefont {J.}~\bibnamefont {Baugh}},\ }\href
	{\doibase 10.1103/PhysRevB.94.155417} {\bibfield  {journal} {\bibinfo
			{journal} {Phys. Rev. B}\ }\textbf {\bibinfo {volume} {94}},\ \bibinfo
		{pages} {155417} (\bibinfo {year} {2016})}\BibitemShut {NoStop}%
	\bibitem [{\citenamefont {Silva}\ and\ \citenamefont
		{Vernek}(2016)}]{Silva_2016}%
	\BibitemOpen
	\bibfield  {author} {\bibinfo {author} {\bibfnamefont {J.~F.}\ \bibnamefont
			{Silva}}\ and\ \bibinfo {author} {\bibfnamefont {E.}~\bibnamefont {Vernek}},\
	}\href {\doibase 10.1088/0953-8984/28/43/435702} {\bibfield  {journal}
		{\bibinfo  {journal} {Journal of Physics: Condensed Matter}\ }\textbf
		{\bibinfo {volume} {28}},\ \bibinfo {pages} {435702} (\bibinfo {year}
		{2016})}\BibitemShut {NoStop}%
	\bibitem [{\citenamefont {Ramos-Andrade}\ \emph {et~al.}(2018)\citenamefont
		{Ramos-Andrade}, \citenamefont {Orellana},\ and\ \citenamefont
		{Ulloa}}]{Ramos_Andrade_2018}%
	\BibitemOpen
	\bibfield  {author} {\bibinfo {author} {\bibfnamefont {J.~P.}\ \bibnamefont
			{Ramos-Andrade}}, \bibinfo {author} {\bibfnamefont {P.~A.}\ \bibnamefont
			{Orellana}}, \ and\ \bibinfo {author} {\bibfnamefont {S.~E.}\ \bibnamefont
			{Ulloa}},\ }\href {\doibase 10.1088/1361-648x/aaa1b2} {\bibfield  {journal}
		{\bibinfo  {journal} {Journal of Physics: Condensed Matter}\ }\textbf
		{\bibinfo {volume} {30}},\ \bibinfo {pages} {045301} (\bibinfo {year}
		{2018})}\BibitemShut {NoStop}%
	\bibitem [{\citenamefont {Vernek}\ \emph {et~al.}(2014)\citenamefont {Vernek},
		\citenamefont {Penteado}, \citenamefont {Seridonio},\ and\ \citenamefont
		{Egues}}]{vernek1}%
	\BibitemOpen
	\bibfield  {author} {\bibinfo {author} {\bibfnamefont {E.}~\bibnamefont
			{Vernek}}, \bibinfo {author} {\bibfnamefont {P.~H.}\ \bibnamefont
			{Penteado}}, \bibinfo {author} {\bibfnamefont {A.~C.}\ \bibnamefont
			{Seridonio}}, \ and\ \bibinfo {author} {\bibfnamefont {J.~C.}\ \bibnamefont
			{Egues}},\ }\href {\doibase 10.1103/PhysRevB.89.165314} {\bibfield  {journal}
		{\bibinfo  {journal} {Physical Review B}\ }\textbf {\bibinfo {volume} {89}},\
		\bibinfo {pages} {165314} (\bibinfo {year} {2014})}\BibitemShut {NoStop}%
	\bibitem [{\citenamefont {Deng}\ \emph {et~al.}(2018)\citenamefont {Deng},
		\citenamefont {Vaitiek\ifmmode~\dot{e}\else \.{e}\fi{}nas}, \citenamefont
		{Prada}, \citenamefont {San-Jose}, \citenamefont {Nyg\aa{}rd}, \citenamefont
		{Krogstrup}, \citenamefont {Aguado},\ and\ \citenamefont
		{Marcus}}]{Deng:Phys.Rev.B:98:085125:2018}%
	\BibitemOpen
	\bibfield  {author} {\bibinfo {author} {\bibfnamefont {M.-T.}\ \bibnamefont
			{Deng}}, \bibinfo {author} {\bibfnamefont {S.}~\bibnamefont
			{Vaitiek\ifmmode~\dot{e}\else \.{e}\fi{}nas}}, \bibinfo {author}
		{\bibfnamefont {E.}~\bibnamefont {Prada}}, \bibinfo {author} {\bibfnamefont
			{P.}~\bibnamefont {San-Jose}}, \bibinfo {author} {\bibfnamefont
			{J.}~\bibnamefont {Nyg\aa{}rd}}, \bibinfo {author} {\bibfnamefont
			{P.}~\bibnamefont {Krogstrup}}, \bibinfo {author} {\bibfnamefont
			{R.}~\bibnamefont {Aguado}}, \ and\ \bibinfo {author} {\bibfnamefont {C.~M.}\
			\bibnamefont {Marcus}},\ }\href {\doibase 10.1103/PhysRevB.98.085125}
	{\bibfield  {journal} {\bibinfo  {journal} {Phys. Rev. B}\ }\textbf {\bibinfo
			{volume} {98}},\ \bibinfo {pages} {085125} (\bibinfo {year}
		{2018})}\BibitemShut {NoStop}%
	\bibitem [{\citenamefont {Hewson}(1993)}]{Hewson}%
	\BibitemOpen
	\bibfield  {author} {\bibinfo {author} {\bibfnamefont {A.~C.}\ \bibnamefont
			{Hewson}},\ }\href@noop {} {\emph {\bibinfo {title} {The Kondo problem to
				heavy fermions}}}\ (\bibinfo  {publisher} {Cambridge University Press},\
	\bibinfo {year} {1993})\BibitemShut {NoStop}%
	\bibitem [{\citenamefont {Zhang}\ \emph {et~al.}(2013)\citenamefont {Zhang},
		\citenamefont {Kane},\ and\ \citenamefont {Mele}}]{Mele}%
	\BibitemOpen
	\bibfield  {author} {\bibinfo {author} {\bibfnamefont {F.}~\bibnamefont
			{Zhang}}, \bibinfo {author} {\bibfnamefont {C.~L.}\ \bibnamefont {Kane}}, \
		and\ \bibinfo {author} {\bibfnamefont {E.~J.}\ \bibnamefont {Mele}},\ }\href
	{\doibase 10.1103/PhysRevLett.111.056402} {\bibfield  {journal} {\bibinfo
			{journal} {Phys. Rev. Lett.}\ }\textbf {\bibinfo {volume} {111}},\ \bibinfo
		{pages} {056402} (\bibinfo {year} {2013})}\BibitemShut {NoStop}%
	\bibitem [{\citenamefont {Jiang}\ \emph {et~al.}(2019)\citenamefont {Jiang},
		\citenamefont {Dai},\ and\ \citenamefont {Wang}}]{Wang}%
	\BibitemOpen
	\bibfield  {author} {\bibinfo {author} {\bibfnamefont {K.}~\bibnamefont
			{Jiang}}, \bibinfo {author} {\bibfnamefont {X.}~\bibnamefont {Dai}}, \ and\
		\bibinfo {author} {\bibfnamefont {Z.}~\bibnamefont {Wang}},\ }\href {\doibase
		10.1103/PhysRevX.9.011033} {\bibfield  {journal} {\bibinfo  {journal} {Phys.
				Rev. X}\ }\textbf {\bibinfo {volume} {9}},\ \bibinfo {pages} {011033}
		(\bibinfo {year} {2019})}\BibitemShut {NoStop}%
	\bibitem [{\citenamefont {Reynoso}\ and\ \citenamefont
		{Frustaglia}(2013)}]{Frustaglia}%
	\BibitemOpen
	\bibfield  {author} {\bibinfo {author} {\bibfnamefont {A.~A.}\ \bibnamefont
			{Reynoso}}\ and\ \bibinfo {author} {\bibfnamefont {D.}~\bibnamefont
			{Frustaglia}},\ }\href {\doibase 10.1103/PhysRevB.87.115420} {\bibfield
		{journal} {\bibinfo  {journal} {Phys. Rev. B}\ }\textbf {\bibinfo {volume}
			{87}},\ \bibinfo {pages} {115420} (\bibinfo {year} {2013})}\BibitemShut
	{NoStop}%
	\bibitem [{\citenamefont {Sato}\ and\ \citenamefont
		{Fujimoto}(2009)}]{Fugimoto}%
	\BibitemOpen
	\bibfield  {author} {\bibinfo {author} {\bibfnamefont {M.}~\bibnamefont
			{Sato}}\ and\ \bibinfo {author} {\bibfnamefont {S.}~\bibnamefont
			{Fujimoto}},\ }\href {\doibase 10.1103/PhysRevB.79.094504} {\bibfield
		{journal} {\bibinfo  {journal} {Phys. Rev. B}\ }\textbf {\bibinfo {volume}
			{79}},\ \bibinfo {pages} {094504} (\bibinfo {year} {2009})}\BibitemShut
	{NoStop}%
	\bibitem [{\citenamefont {Pillet}\ \emph {et~al.}(2013)\citenamefont {Pillet},
		\citenamefont {Joyez}, \citenamefont {\ifmmode~\check{Z}\else
			\v{Z}\fi{}itko},\ and\ \citenamefont {Goffman}}]{Pillet}%
	\BibitemOpen
	\bibfield  {author} {\bibinfo {author} {\bibfnamefont {J.-D.}\ \bibnamefont
			{Pillet}}, \bibinfo {author} {\bibfnamefont {P.}~\bibnamefont {Joyez}},
		\bibinfo {author} {\bibfnamefont {R.}~\bibnamefont {\ifmmode~\check{Z}\else
				\v{Z}\fi{}itko}}, \ and\ \bibinfo {author} {\bibfnamefont {M.~F.}\
			\bibnamefont {Goffman}},\ }\href {\doibase 10.1103/PhysRevB.88.045101}
	{\bibfield  {journal} {\bibinfo  {journal} {Phys. Rev. B}\ }\textbf {\bibinfo
			{volume} {88}},\ \bibinfo {pages} {045101} (\bibinfo {year}
		{2013})}\BibitemShut {NoStop}%
	\bibitem [{\citenamefont {Cheng}\ \emph {et~al.}(2014)\citenamefont {Cheng},
		\citenamefont {Becker}, \citenamefont {Bauer},\ and\ \citenamefont
		{Lutchyn}}]{PhysRevX.4.031051}%
	\BibitemOpen
	\bibfield  {author} {\bibinfo {author} {\bibfnamefont {M.}~\bibnamefont
			{Cheng}}, \bibinfo {author} {\bibfnamefont {M.}~\bibnamefont {Becker}},
		\bibinfo {author} {\bibfnamefont {B.}~\bibnamefont {Bauer}}, \ and\ \bibinfo
		{author} {\bibfnamefont {R.~M.}\ \bibnamefont {Lutchyn}},\ }\href {\doibase
		10.1103/PhysRevX.4.031051} {\bibfield  {journal} {\bibinfo  {journal} {Phys.
				Rev. X}\ }\textbf {\bibinfo {volume} {4}},\ \bibinfo {pages} {031051}
		(\bibinfo {year} {2014})}\BibitemShut {NoStop}%
	\bibitem [{\citenamefont {Lee}\ \emph {et~al.}(2013)\citenamefont {Lee},
		\citenamefont {Lim},\ and\ \citenamefont {L\'opez}}]{Lee}%
	\BibitemOpen
	\bibfield  {author} {\bibinfo {author} {\bibfnamefont {M.}~\bibnamefont
			{Lee}}, \bibinfo {author} {\bibfnamefont {J.~S.}\ \bibnamefont {Lim}}, \ and\
		\bibinfo {author} {\bibfnamefont {R.}~\bibnamefont {L\'opez}},\ }\href
	{\doibase 10.1103/PhysRevB.87.241402} {\bibfield  {journal} {\bibinfo
			{journal} {Phys. Rev. B}\ }\textbf {\bibinfo {volume} {87}},\ \bibinfo
		{pages} {241402} (\bibinfo {year} {2013})}\BibitemShut {NoStop}%
	\bibitem [{\citenamefont {Chirla}\ \emph {et~al.}(2014)\citenamefont {Chirla},
		\citenamefont {Dinu}, \citenamefont {Moldoveanu},\ and\ \citenamefont
		{Moca}}]{Chirla:Phys.Rev.B:90:195108:2014}%
	\BibitemOpen
	\bibfield  {author} {\bibinfo {author} {\bibfnamefont {R.}~\bibnamefont
			{Chirla}}, \bibinfo {author} {\bibfnamefont {I.~V.}\ \bibnamefont {Dinu}},
		\bibinfo {author} {\bibfnamefont {V.}~\bibnamefont {Moldoveanu}}, \ and\
		\bibinfo {author} {\bibfnamefont {C.~u. u. u. u. P. m.~c.}\ \bibnamefont
			{Moca}},\ }\href {\doibase 10.1103/PhysRevB.90.195108} {\bibfield  {journal}
		{\bibinfo  {journal} {Phys. Rev. B}\ }\textbf {\bibinfo {volume} {90}},\
		\bibinfo {pages} {195108} (\bibinfo {year} {2014})}\BibitemShut {NoStop}%
	\bibitem [{\citenamefont {Ruiz-Tijerina}\ \emph {et~al.}(2015)\citenamefont
		{Ruiz-Tijerina}, \citenamefont {Vernek}, \citenamefont {Dias~da Silva},\ and\
		\citenamefont {Egues}}]{Tijerina}%
	\BibitemOpen
	\bibfield  {author} {\bibinfo {author} {\bibfnamefont {D.~A.}\ \bibnamefont
			{Ruiz-Tijerina}}, \bibinfo {author} {\bibfnamefont {E.}~\bibnamefont
			{Vernek}}, \bibinfo {author} {\bibfnamefont {L.~G. G.~V.}\ \bibnamefont
			{Dias~da Silva}}, \ and\ \bibinfo {author} {\bibfnamefont {J.~C.}\
			\bibnamefont {Egues}},\ }\href {\doibase 10.1103/PhysRevB.91.115435}
	{\bibfield  {journal} {\bibinfo  {journal} {Phys. Rev. B}\ }\textbf {\bibinfo
			{volume} {91}},\ \bibinfo {pages} {115435} (\bibinfo {year}
		{2015})}\BibitemShut {NoStop}%
	\bibitem [{\citenamefont {G{\'{o}}rski}\ \emph {et~al.}(2018)\citenamefont
		{G{\'{o}}rski}, \citenamefont {Bara{\'{n}}ski}, \citenamefont {Weymann},\
		and\ \citenamefont {Doma{\'{n}}ski}}]{Gorski2018}%
	\BibitemOpen
	\bibfield  {author} {\bibinfo {author} {\bibfnamefont {G.}~\bibnamefont
			{G{\'{o}}rski}}, \bibinfo {author} {\bibfnamefont {J.}~\bibnamefont
			{Bara{\'{n}}ski}}, \bibinfo {author} {\bibfnamefont {I.}~\bibnamefont
			{Weymann}}, \ and\ \bibinfo {author} {\bibfnamefont {T.}~\bibnamefont
			{Doma{\'{n}}ski}},\ }\href {\doibase 10.1038/s41598-018-33529-1} {\bibfield
		{journal} {\bibinfo  {journal} {Scientific Reports}\ }\textbf {\bibinfo
			{volume} {8}},\ \bibinfo {pages} {1} (\bibinfo {year} {2018})}\BibitemShut
	{NoStop}%
	\bibitem [{\citenamefont {Weymann}\ and\ \citenamefont
		{W\'ojcik}(2017)}]{PhysRevB.95.155427}%
	\BibitemOpen
	\bibfield  {author} {\bibinfo {author} {\bibfnamefont {I.}~\bibnamefont
			{Weymann}}\ and\ \bibinfo {author} {\bibfnamefont {K.~P.}\ \bibnamefont
			{W\'ojcik}},\ }\href {\doibase 10.1103/PhysRevB.95.155427} {\bibfield
		{journal} {\bibinfo  {journal} {Phys. Rev. B}\ }\textbf {\bibinfo {volume}
			{95}},\ \bibinfo {pages} {155427} (\bibinfo {year} {2017})}\BibitemShut
	{NoStop}%
	\bibitem [{\citenamefont {Cifuentes}\ and\ \citenamefont
		{da~Silva}(2019)}]{Cifuentes}%
	\BibitemOpen
	\bibfield  {author} {\bibinfo {author} {\bibfnamefont {J.~D.}\ \bibnamefont
			{Cifuentes}}\ and\ \bibinfo {author} {\bibfnamefont {L.~G. G. V.~D.}\
			\bibnamefont {da~Silva}},\ }\href {\doibase 10.1103/PhysRevB.100.085429}
	{\bibfield  {journal} {\bibinfo  {journal} {Phys. Rev. B}\ }\textbf {\bibinfo
			{volume} {100}},\ \bibinfo {pages} {085429} (\bibinfo {year}
		{2019})}\BibitemShut {NoStop}%
	\bibitem [{\citenamefont {da~Cruz}(2016)}]{Adonai:Thesis:2016}%
	\BibitemOpen
	\bibfield  {author} {\bibinfo {author} {\bibfnamefont {A.~R.}\ \bibnamefont
			{da~Cruz}},\ }\emph {\bibinfo {title} {{Fusing Majorana modes in
				quantum-dots}}},\ \href@noop {} {Master's thesis},\ \bibinfo  {school}
	{Instituto de F\'isica de S\~ao Carlos}, \bibinfo {address} {Brazil}
	(\bibinfo {year} {2016})\BibitemShut {NoStop}%
	\bibitem [{\citenamefont {Hoffman}\ \emph {et~al.}(2017)\citenamefont
		{Hoffman}, \citenamefont {Chevallier}, \citenamefont {Loss},\ and\
		\citenamefont {Klinovaja}}]{Hoffman:Phys.Rev.B:045440:2017}%
	\BibitemOpen
	\bibfield  {author} {\bibinfo {author} {\bibfnamefont {S.}~\bibnamefont
			{Hoffman}}, \bibinfo {author} {\bibfnamefont {D.}~\bibnamefont {Chevallier}},
		\bibinfo {author} {\bibfnamefont {D.}~\bibnamefont {Loss}}, \ and\ \bibinfo
		{author} {\bibfnamefont {J.}~\bibnamefont {Klinovaja}},\ }\href {\doibase
		10.1103/PhysRevB.96.045440} {\bibfield  {journal} {\bibinfo  {journal} {Phys.
				Rev. B}\ }\textbf {\bibinfo {volume} {96}},\ \bibinfo {pages} {045440}
		(\bibinfo {year} {2017})}\BibitemShut {NoStop}%
	\bibitem [{\citenamefont {Prada}\ \emph {et~al.}(2017)\citenamefont {Prada},
		\citenamefont {Aguado},\ and\ \citenamefont
		{San-Jose}}]{Prada:Phys.Rev.B:96:085418:2017}%
	\BibitemOpen
	\bibfield  {author} {\bibinfo {author} {\bibfnamefont {E.}~\bibnamefont
			{Prada}}, \bibinfo {author} {\bibfnamefont {R.}~\bibnamefont {Aguado}}, \
		and\ \bibinfo {author} {\bibfnamefont {P.}~\bibnamefont {San-Jose}},\ }\href
	{\doibase 10.1103/PhysRevB.96.085418} {\bibfield  {journal} {\bibinfo
			{journal} {Phys. Rev. B}\ }\textbf {\bibinfo {volume} {96}},\ \bibinfo
		{pages} {085418} (\bibinfo {year} {2017})}\BibitemShut {NoStop}%
	\bibitem [{\citenamefont {Emery}\ and\ \citenamefont
		{Kivelson}(1992)}]{PhysRevB.46.10812}%
	\BibitemOpen
	\bibfield  {author} {\bibinfo {author} {\bibfnamefont {V.~J.}\ \bibnamefont
			{Emery}}\ and\ \bibinfo {author} {\bibfnamefont {S.}~\bibnamefont
			{Kivelson}},\ }\href {\doibase 10.1103/PhysRevB.46.10812} {\bibfield
		{journal} {\bibinfo  {journal} {Phys. Rev. B}\ }\textbf {\bibinfo {volume}
			{46}},\ \bibinfo {pages} {10812} (\bibinfo {year} {1992})}\BibitemShut
	{NoStop}%
	\bibitem [{\citenamefont {Coleman}\ \emph {et~al.}(1995)\citenamefont
		{Coleman}, \citenamefont {Ioffe},\ and\ \citenamefont
		{Tsvelik}}]{PhysRevB.52.6611}%
	\BibitemOpen
	\bibfield  {author} {\bibinfo {author} {\bibfnamefont {P.}~\bibnamefont
			{Coleman}}, \bibinfo {author} {\bibfnamefont {L.~B.}\ \bibnamefont {Ioffe}},
		\ and\ \bibinfo {author} {\bibfnamefont {A.~M.}\ \bibnamefont {Tsvelik}},\
	}\href {\doibase 10.1103/PhysRevB.52.6611} {\bibfield  {journal} {\bibinfo
			{journal} {Phys. Rev. B}\ }\textbf {\bibinfo {volume} {52}},\ \bibinfo
		{pages} {6611} (\bibinfo {year} {1995})}\BibitemShut {NoStop}%
	\bibitem [{\citenamefont {Cox}\ and\ \citenamefont {Zawadowski}(1998)}]{Cox}%
	\BibitemOpen
	\bibfield  {author} {\bibinfo {author} {\bibfnamefont {D.~L.}\ \bibnamefont
			{Cox}}\ and\ \bibinfo {author} {\bibfnamefont {A.}~\bibnamefont
			{Zawadowski}},\ }\href {\doibase 10.1080/000187398243500} {\bibfield
		{journal} {\bibinfo  {journal} {Advances in Physics}\ }\textbf {\bibinfo
			{volume} {47}},\ \bibinfo {pages} {599} (\bibinfo {year} {1998})}\BibitemShut
	{NoStop}%
	\bibitem [{\citenamefont {Zhang}\ \emph {et~al.}(1999)\citenamefont {Zhang},
		\citenamefont {Hewson},\ and\ \citenamefont {Bulla}}]{ZHANG}%
	\BibitemOpen
	\bibfield  {author} {\bibinfo {author} {\bibfnamefont {G.-M.}\ \bibnamefont
			{Zhang}}, \bibinfo {author} {\bibfnamefont {A.}~\bibnamefont {Hewson}}, \
		and\ \bibinfo {author} {\bibfnamefont {R.}~\bibnamefont {Bulla}},\ }\href
	{\doibase https://doi.org/10.1016/S0038-1098(99)00294-X} {\bibfield
		{journal} {\bibinfo  {journal} {Solid State Communications}\ }\textbf
		{\bibinfo {volume} {112}},\ \bibinfo {pages} {105 } (\bibinfo {year}
		{1999})}\BibitemShut {NoStop}%
	\bibitem [{\citenamefont {Smirnov}(2015)}]{Smirnov:Phys.Rev.B:92:195312:2015}%
	\BibitemOpen
	\bibfield  {author} {\bibinfo {author} {\bibfnamefont {S.}~\bibnamefont
			{Smirnov}},\ }\href {\doibase 10.1103/PhysRevB.92.195312} {\bibfield
		{journal} {\bibinfo  {journal} {Phys. Rev. B}\ }\textbf {\bibinfo {volume}
			{92}},\ \bibinfo {pages} {195312} (\bibinfo {year} {2015})}\BibitemShut
	{NoStop}%
	\bibitem [{\citenamefont {Sela}\ \emph {et~al.}(2019)\citenamefont {Sela},
		\citenamefont {Oreg}, \citenamefont {Plugge}, \citenamefont {Hartman},
		\citenamefont {L\"uscher},\ and\ \citenamefont
		{Folk}}]{PhysRevLett.123.147702}%
	\BibitemOpen
	\bibfield  {author} {\bibinfo {author} {\bibfnamefont {E.}~\bibnamefont
			{Sela}}, \bibinfo {author} {\bibfnamefont {Y.}~\bibnamefont {Oreg}}, \bibinfo
		{author} {\bibfnamefont {S.}~\bibnamefont {Plugge}}, \bibinfo {author}
		{\bibfnamefont {N.}~\bibnamefont {Hartman}}, \bibinfo {author} {\bibfnamefont
			{S.}~\bibnamefont {L\"uscher}}, \ and\ \bibinfo {author} {\bibfnamefont
			{J.}~\bibnamefont {Folk}},\ }\href {\doibase 10.1103/PhysRevLett.123.147702}
	{\bibfield  {journal} {\bibinfo  {journal} {Phys. Rev. Lett.}\ }\textbf
		{\bibinfo {volume} {123}},\ \bibinfo {pages} {147702} (\bibinfo {year}
		{2019})}\BibitemShut {NoStop}%
	\bibitem [{\citenamefont {{Nozi\`eres, Ph.}}\ and\ \citenamefont {{Blandin,
				A.}}(1980)}]{Nozieres}%
	\BibitemOpen
	\bibfield  {author} {\bibinfo {author} {\bibnamefont {{Nozi\`eres, Ph.}}}\
		and\ \bibinfo {author} {\bibnamefont {{Blandin, A.}}},\ }\href {\doibase
		10.1051/jphys:01980004103019300} {\bibfield  {journal} {\bibinfo  {journal}
			{J. Phys. France}\ }\textbf {\bibinfo {volume} {41}},\ \bibinfo {pages} {193}
		(\bibinfo {year} {1980})}\BibitemShut {NoStop}%
	\bibitem [{\citenamefont {Affleck}\ and\ \citenamefont
		{Ludwig}(1991{\natexlab{a}})}]{PhysRevLett.67.161}%
	\BibitemOpen
	\bibfield  {author} {\bibinfo {author} {\bibfnamefont {I.}~\bibnamefont
			{Affleck}}\ and\ \bibinfo {author} {\bibfnamefont {A.~W.~W.}\ \bibnamefont
			{Ludwig}},\ }\href {\doibase 10.1103/PhysRevLett.67.161} {\bibfield
		{journal} {\bibinfo  {journal} {Phys. Rev. Lett.}\ }\textbf {\bibinfo
			{volume} {67}},\ \bibinfo {pages} {161} (\bibinfo {year}
		{1991}{\natexlab{a}})}\BibitemShut {NoStop}%
	\bibitem [{\citenamefont {Hoffman}\ \emph {et~al.}(2016)\citenamefont
		{Hoffman}, \citenamefont {Schrade}, \citenamefont {Klinovaja},\ and\
		\citenamefont {Loss}}]{PhysRevB.94.045316}%
	\BibitemOpen
	\bibfield  {author} {\bibinfo {author} {\bibfnamefont {S.}~\bibnamefont
			{Hoffman}}, \bibinfo {author} {\bibfnamefont {C.}~\bibnamefont {Schrade}},
		\bibinfo {author} {\bibfnamefont {J.}~\bibnamefont {Klinovaja}}, \ and\
		\bibinfo {author} {\bibfnamefont {D.}~\bibnamefont {Loss}},\ }\href {\doibase
		10.1103/PhysRevB.94.045316} {\bibfield  {journal} {\bibinfo  {journal} {Phys.
				Rev. B}\ }\textbf {\bibinfo {volume} {94}},\ \bibinfo {pages} {045316}
		(\bibinfo {year} {2016})}\BibitemShut {NoStop}%
	\bibitem [{\citenamefont {Anderson}(1970)}]{Anderson1970}%
	\BibitemOpen
	\bibfield  {author} {\bibinfo {author} {\bibfnamefont {P.~W.}\ \bibnamefont
			{Anderson}},\ }\href {\doibase 10.1088/0022-3719/3/12/008} {\bibfield
		{journal} {\bibinfo  {journal} {J. Phys. C: Sol. St. Phys.}\ }\textbf
		{\bibinfo {volume} {3}},\ \bibinfo {pages} {2436} (\bibinfo {year}
		{1970})}\BibitemShut {NoStop}%
	\bibitem [{\citenamefont {Zubarev}(1960)}]{Sov.Phys.Usp..3.320}%
	\BibitemOpen
	\bibfield  {author} {\bibinfo {author} {\bibfnamefont {D.~N.}\ \bibnamefont
			{Zubarev}},\ }\href@noop {} {\bibfield  {journal} {\bibinfo  {journal} {Sov.
				Phys. Usp.}\ }\textbf {\bibinfo {volume} {3}},\ \bibinfo {pages} {320}
		(\bibinfo {year} {1960})}\BibitemShut {NoStop}%
	\bibitem [{\citenamefont {Tsvelick}(1985)}]{Tsvelick_1985}%
	\BibitemOpen
	\bibfield  {author} {\bibinfo {author} {\bibfnamefont {A.~M.}\ \bibnamefont
			{Tsvelick}},\ }\href {\doibase 10.1088/0022-3719/18/1/020} {\bibfield
		{journal} {\bibinfo  {journal} {Journal of Physics C: Solid State Physics}\
		}\textbf {\bibinfo {volume} {18}},\ \bibinfo {pages} {159} (\bibinfo {year}
		{1985})}\BibitemShut {NoStop}%
	\bibitem [{\citenamefont {Affleck}\ and\ \citenamefont
		{Ludwig}(1991{\natexlab{b}})}]{Lud1}%
	\BibitemOpen
	\bibfield  {author} {\bibinfo {author} {\bibfnamefont {I.}~\bibnamefont
			{Affleck}}\ and\ \bibinfo {author} {\bibfnamefont {A.~W.}\ \bibnamefont
			{Ludwig}},\ }\href {\doibase https://doi.org/10.1016/0550-3213(91)90419-X}
	{\bibfield  {journal} {\bibinfo  {journal} {Nuclear Physics B}\ }\textbf
		{\bibinfo {volume} {360}},\ \bibinfo {pages} {641 } (\bibinfo {year}
		{1991}{\natexlab{b}})}\BibitemShut {NoStop}%
	\bibitem [{\citenamefont {Affleck}\ and\ \citenamefont
		{Ludwig}(1993)}]{PhysRevB.48.7297}%
	\BibitemOpen
	\bibfield  {author} {\bibinfo {author} {\bibfnamefont {I.}~\bibnamefont
			{Affleck}}\ and\ \bibinfo {author} {\bibfnamefont {A.~W.~W.}\ \bibnamefont
			{Ludwig}},\ }\href {\doibase 10.1103/PhysRevB.48.7297} {\bibfield  {journal}
		{\bibinfo  {journal} {Phys. Rev. B}\ }\textbf {\bibinfo {volume} {48}},\
		\bibinfo {pages} {7297} (\bibinfo {year} {1993})}\BibitemShut {NoStop}%
	\bibitem [{\citenamefont {Gan}\ \emph {et~al.}(1993)\citenamefont {Gan},
		\citenamefont {Andrei},\ and\ \citenamefont {Coleman}}]{PhysRevLett.70.686}%
	\BibitemOpen
	\bibfield  {author} {\bibinfo {author} {\bibfnamefont {J.}~\bibnamefont
			{Gan}}, \bibinfo {author} {\bibfnamefont {N.}~\bibnamefont {Andrei}}, \ and\
		\bibinfo {author} {\bibfnamefont {P.}~\bibnamefont {Coleman}},\ }\href
	{\doibase 10.1103/PhysRevLett.70.686} {\bibfield  {journal} {\bibinfo
			{journal} {Phys. Rev. Lett.}\ }\textbf {\bibinfo {volume} {70}},\ \bibinfo
		{pages} {686} (\bibinfo {year} {1993})}\BibitemShut {NoStop}%
	\bibitem [{\citenamefont {Fabrizio}\ and\ \citenamefont
		{Gogolin}(1994)}]{PhysRevB.50.17732}%
	\BibitemOpen
	\bibfield  {author} {\bibinfo {author} {\bibfnamefont {M.}~\bibnamefont
			{Fabrizio}}\ and\ \bibinfo {author} {\bibfnamefont {A.~O.}\ \bibnamefont
			{Gogolin}},\ }\href {\doibase 10.1103/PhysRevB.50.17732} {\bibfield
		{journal} {\bibinfo  {journal} {Phys. Rev. B}\ }\textbf {\bibinfo {volume}
			{50}},\ \bibinfo {pages} {17732} (\bibinfo {year} {1994})}\BibitemShut
	{NoStop}%
\end{thebibliography}
\end{document}